\documentclass[lettersize,journal]{IEEEtran}
\usepackage{amsmath,amsfonts}
\usepackage{algorithmic}
\usepackage{algorithm}
\usepackage{array}
\usepackage{textcomp}
\usepackage{stfloats}
\usepackage{url}
\usepackage{verbatim}
\usepackage{graphicx}
\usepackage{cite}
\usepackage{makecell}
\usepackage{graphicx}
\usepackage{bm}
\usepackage{amssymb}
\usepackage{amsthm}
\usepackage[caption=false,font=footnotesize,labelfont=rm,textfont=rm]{subfig}
\newtheorem{thm}{Theorem}

\begin{document}

\title{Analog Self-Interference Cancellation in Full-Duplex Radios: A Fundamental Limit Perspective}

\author{Limin Liao, Jun Sun, Junzhi Wang, Yingzhuang Liu
\thanks{Limin Liao, Jun Sun, Junzhi Wang, Yingzhuang Liu are with the School of Electronic Information and Communications, Huazhong University of Science and Technology, Wuhan, China. (e-mail: liaolimin001@hust.edu.cn; juns@hust.edu.cn; wangjunzhi@hust.edu.cn; liuyz@hust.edu.cn)}
}



\maketitle

\begin{abstract}
Analog self-interference cancellation (A-SIC) plays a crucial role in the implementation of in-band full-duplex (IBFD) radios, due to the fact that the inherent transmit (Tx) noise can only be addressed in the analog domain. It is thus natural to ask what the performance limit of A-SIC is in practical systems, which is still quite underexplored so far. In this paper, we aim to close this gap by characterizing the fundamental performance of A-SIC which employs the common multi-tap delay (MTD) architecture, by accounting for the following practical issues: 1) Nonstationarity of the Tx signal; 2) Nonlinear distortions on the Tx signal; 3) Multipath channel corresponding to the self-interference (SI); 4) Maximum amplitude constraint on the MTD tap weights. Our findings include: 1) The average approximation error for the cyclostationary Tx signals is \textit{equal to} that for the \textit{stationary} white \textit{Gaussian} process, thus greatly simplifying the performance analysis and the optimization procedure. 2) The approximation error for the multipath SI channel can be \textit{decomposed} as the sum of the approximation error for the single-path scenario. By leveraging these structural results, the optimization framework and algorithms which characterize the fundamental limit of A-SIC, by taking into account all the aforementioned practical factors, are provided.
\end{abstract}


\bibliographystyle{plain}

\section{Introduction}
\IEEEPARstart{I}{n}-band full-duplex (IBFD) technology has the potential to double the spectrum efficiency of wireless systems by simultaneous transmission and reception over the same frequency band, which is highly attractive as the spectrum becomes increasingly scarce nowadays \cite{hong2014applications}. 

To exploit the potential gain of IBFD, the biggest challenge lies in canceling the extremely strong self-interference (SI) due to the super-short distance between the transmit (Tx) antenna to the receive (Rx) chain. Taking the Wi-Fi system as an example, the Tx power of an IBFD radio is limited to about 20 dBm, while the Rx noise floor is around -90 dBm for 80 MHz of bandwidth. In order to receive the desired signal with negligible degradation for the Rx chain, we have to reduce the power of SI to the level of noise floor or below. In other words, the self-interference cancellation (SIC) ability of IBFD radios have to be no less than 110 dB, which is obviously a very big challenge for the implementation. 

To cope with the extremely strong SI of IBFD, we need to perform SIC in both the analog and digital domain. By combining SIC in these two domains, extensive efforts have been made to boost the overall SIC ability, in hope of meeting the above requirement. Notable works include: \cite{bharadia2013full} demonstrated a Wi-Fi IBFD radio achieving 110 dB of SIC ability, with 63 dB of cancellation by analog SIC (aided by 15 dB isolation via the circulator) and 47 dB by digital SIC (D-SIC).  \cite{korpi2016compact} employed a different version of analog SIC (A-SIC), i.e., the closed-loop (adaptive) minimum mean square error (MMSE) for A-SIC, and reported over 100 dB of SIC ability, by incorporating a high-isolation (65 dB) antenna. Moreover, an architecture featuring a digital-assisted A-SIC combined with a 4-tap delay A-SIC achieved 107 dB of SIC ability with 20 MHz of bandwidth\cite{cheng2023digital}. Despite these progresses, unfortunately, our understanding on the fundamental limit of SIC is still very limited, especially when considering practical issues, for example, the cyclostationarity of the Tx signal, the multipath nature of the SI channel and the amplitude constraint on tap weights. In this paper, our aim is to close the gap by exploring the fundamental limit of A-SIC. And in its sequel, we will address its counterpart of D-SIC.

A-SIC is of critical importance for the following two reasons: 1) the inevitable Tx noise is \textit{impossible} to be cancelled in the digital domain; 2) the quantization noise due to the limited resolution of the analog-to-digital converter (ADC) poses an upper limit on the performance of SIC in the digital domain. Consequently, the SI has to be reduced to a rather low level in the analog domain; otherwise, it would be the bottleneck of the overall SIC performance. Together, it is crucial to explore the fundamental limit of A-SIC.

\IEEEpubidadjcol

The typical architecture of A-SIC, i.e. multi-tap delay (MTD), which is proposed in \cite{bharadia2013full}, is motivated by the Shannon-Nyquist sampling theorem, and its main idea is to reconstruct SI by a linear combination of delayed copies of the Tx signal. Other A-SIC architectures, such as adaptive MMSE in \cite{korpi2016compact} can also be seen in this category. For this reason, our investigation will focus on the MTD architecture. 

There are many existing works on A-SIC; however, most of them are on the architectural and algorithmic side\cite{lee20232,liu2017full,le2019beam,le2020beam,phungamngern2013digital,tamminen2016digitally,alexandropoulos2020full,huang2017radio,kolodziej2016multitap}. Theoretical studies are comparatively much less. In this regard, \cite{choi2013simultaneous} provided a closed-form expression for the residual power of SI based on the MTD A-SIC, assuming the \textit{stationarity} of the Tx signal in \textit{single}-path. 
Taking into account the cyclostationarity of orthogonal frequency division multiplexing (OFDM) signals, \cite{huang2021alms,huang2017radio} analyzed the residual error of A-SIC based on the analog least mean squares (LMS) loop. In addition, considering the non-ideality of the radio frequency (RF) components, \cite{syrjala2014analysis} provided a theoretical analysis on the effects of phase noise introduced by the local oscillator (LO), while the effect of limited accuracy of analog devices, such as attenuators, is analyzed in \cite{zhao2016impacts,wang2016effect}.  

It is worth noting that a systematic study of the performance limit of A-SIC is still lacking, especially in practical scenarios, namely, by taking into account \textit{all} the practical issues: 1) cyclostationarity (rather than stationarity) of the modulated signal; 2) nonlinear distortion of the Tx signal; 3) multipath (rather than single-path) nature of the SI channel; 4) constraint on maximum amplitude of tap weights. Our work aims to close this gap by answering the following questions:

What is the performance limit of A-SIC taking into account all the above-mentioned practical issues? Furthermore, how should we design the optimal MTD (especially the tap delays) of the A-SIC? Our work aims to address these problems in a systematic way. Specifically, our contribution can be summarized as follows:

\begin{itemize}
    \item We prove that when the Tx signal exhibits cyclostationarity, the corresponding MTD A-SIC is \textit{equivalent} to a Wiener filter for stationary Gaussian white noise with limited bandwidth, under the assumption of independent circularly symmetric complex Gaussian symbols. Thus, the performance analysis and subsequent optimization procedures of MTD A-SIC can be greatly simplified.
    \item We demonstrate that in the scenario of multipath SI channels, the approximation error of MTD A-SIC can be tightly bounded as the sum of approximation error associated with each single-path. 
    \item 
    We propose a heuristic and efficient algorithm to determine the tap delays of MTD A-SIC under the constraint of maximum approximation error, which can be seen as the dual problem of characterizing the fundamental limit of MTD A-SIC, i.e., the minimum reconstruction error under the constraint of the total number of taps.
    \item We identify the key factors that determine the fundamental limit of MTD A-SIC, such as tap delays, power-delay profile (PDP), bandwidth, etc.

\end{itemize}

The rest of this paper is organized as follows: In Section \uppercase\expandafter{\romannumeral2}, we provide a brief overview of the background on A-SIC and present the mathematical models for A-SIC. In Section \uppercase\expandafter{\romannumeral3}, we detail the derivation process for the approximation error of SI in multipath channels, along with the optimization algorithms for A-SIC. Additionally, we summarize various factors that influence the performance of A-SIC. In Section \uppercase\expandafter{\romannumeral4}, we compare numerical results of theoretical analysis with experimental simulation results based on a Wi-Fi IBFD radio and evaluate the MTD A-SIC designed by the proposed optimization algorithms against the MTD A-SIC with uniform intervals between adjacent taps as presented in \cite{bharadia2013full}. Finally, the conclusion of this paper is provided in Section \uppercase\expandafter{\romannumeral5}.

\emph{Notations:} In this paper, the circularly symmetric complex Gaussian variable with variance $\sigma^2$ is denoted $\mathcal{CN}(0,\sigma^2)$. The operator $*$ represents the convolution and $\Re \left[  z\right]$ denotes the real part of the complex $z$. $z^*$ is the conjugate of the complex $z$ and $|z|$ is the modulus of $z$. For a matrix $\boldsymbol{X}$, $\boldsymbol{X}^\mathrm{T}$ is the transpose of $\boldsymbol{X}$, $\boldsymbol{X}^\mathrm{H}$ is the conjugate transpose of $\boldsymbol{X}$, and $\boldsymbol{X}^{-1}$ is the inverse of $\boldsymbol{X}$. Particularly, we denote the diagonal matrix $\boldsymbol{A}=[a_{ij}]\in \mathrm{C}^{N\times N}$ by $\mathrm{diag}(a_{11}\ \dots\ a_{NN})$. The Fourier transform for a signal $x(t)$ is denoted as $\mathcal{F}[x(t)]$.

\section{Background and Modeling of A-SIC}
In this section, we first introduce the background of A-SIC and outline the motivation for the theoretical analysis presented. Next, we analyze the advantages as well as disadvantages of two main A-SIC architectures, emphasizing that our theoretical analysis will focus on A-SIC based on MTD as they are capable of canceling Tx noise. Finally, mathematical models of SI and A-SIC are provided.

\subsection{Background and Motivation}
\begin{figure}[!t]
\centering
\includegraphics[width=3.2in]{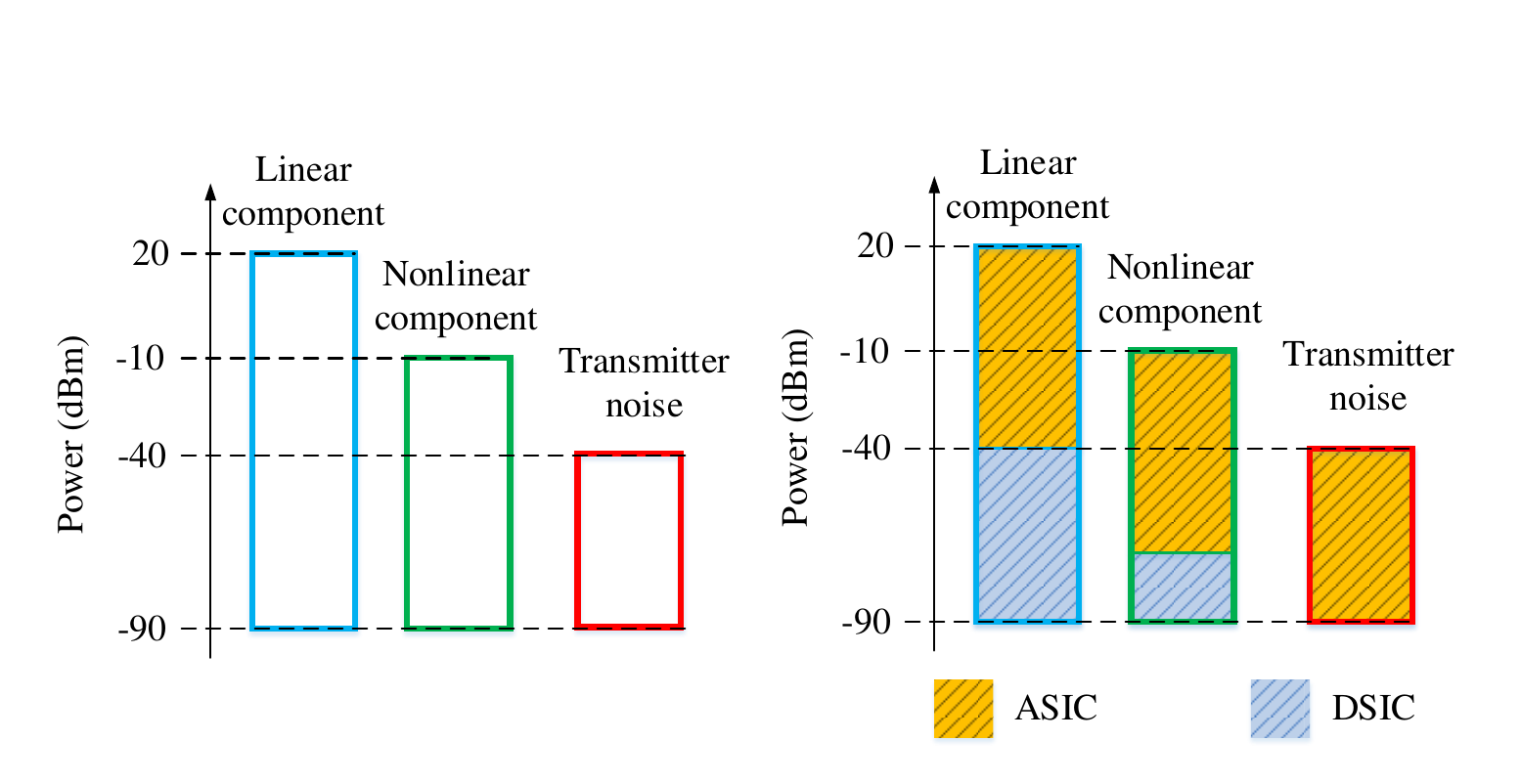}
\caption{left: Power of different components in SI. right: performance requirements for A-SIC and D-SIC.}
\label{fig_1}
\end{figure}

As illustrated in Fig. \ref{fig_1}, the signal radiated by a Wi-Fi radio operating at a Tx power of 20 dBm contains not only linear components but also nonlinear components and Tx noise due to the non-ideality of the RF components \cite{bharadia2013full}. The nonlinear components, introduced due to distortions caused by RF components such as power amplifier (PA), are approximately 30 dB lower than the linear components. Additionally, Tx noise, which includes the inherent thermal noise of the Tx chain and phase noise generated by the LO, is about 60 dB lower than the linear components. Initially, the power of SI must be suppressed to an appropriate level before the low-noise amplifier (LNA) to prevent LNA saturation, requiring the Tx noise to be reduced to the Rx noise floor or below, as it cannot be eliminated in the digital domain. Furthermore, the cancellation performance of D-SIC is constrained by the accuracy of ADCs due to quantization noise. Therefore, A-SIC is a critical and indispensable component of SIC in IBFD radios with high Tx power.  

Generally, A-SIC can be categorized into passive A-SIC and active A-SIC. Passive A-SIC encompasses various antenna technologies, such as circulators \cite{khaledian2018inherent} and separated antennas \cite{korpi2016compact}, which enhance electromagnetic isolation between the Tx and Rx chains to reduce SI. However, passive A-SIC primarily mitigates the direct leakage component of SI, and its performance is influenced by multipath propagation \cite{korpi2016compact}. When the power of multipath components exceeds that of direct leakage after cancellation, passive A-SIC will not help to reduce SI, as the multipath components will dominate the residual SI. In contrast, active A-SIC, which is the primary method for canceling multipath components, reconstructs SI by applying delay, attenuation, phase shift, and other modifications to a replica of the SI obtained from the transmitter. Passive A-SIC is constrained by propagation environments, while D-SIC cannot address Tx noise, even if it can perfectly eliminate linear and nonlinear components in residual SI. Therefore, the ability of active A-SIC to cancel Tx noise emerges as a bottleneck for the overall cancellation performance of SIC, especially in propagation environments with high-power multipath components. Unfortunately, in IBFD radios with higher Tx power, the power of Tx noise tends to increase unless the Tx signal-to-noise ratio (SNR) improves, necessitating A-SIC with superior cancellation performance. These considerations motivate us to explore the performance limits of active A-SIC and its robustness to channel variations.

\subsection{Architecture of A-SIC}
\begin{figure}[!t]
\centering
\includegraphics[width=3.0in]{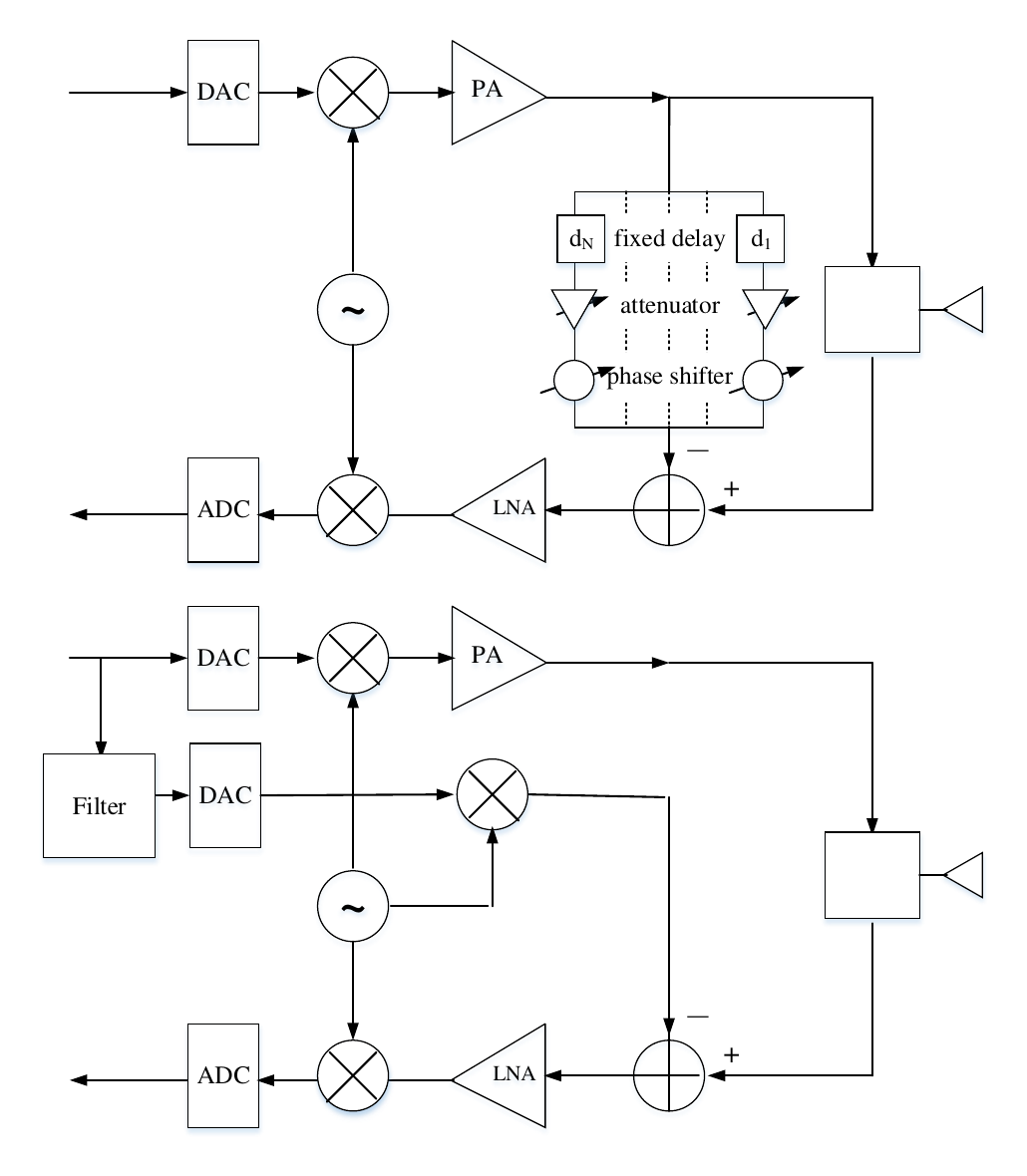}
\caption{The architectures of MTD A-SIC (top) and digital-assisted A-SIC (bottom).}
\label{A-SIC.fig}
\end{figure}

The architectures of active A-SIC can be divided into two categories: 1) Model-based reconstruction, such as digital-assisted A-SIC, which reconstructs various distortions in SI through the actual signal in the digital domain ; 2) Sampling-based reconstruction, such as MTD A-SIC, which reconstructs the SI by sampling the reference signal in the analog domain with different delays and then interpolating them. As illustrated in Fig. \ref{A-SIC.fig}, digital-assisted A-SIC reconstructs distortions, such as in-phase (I/Q) imbalance and PA nonlinearity, associated with SI channels by filtering Tx symbols in the digital domain and then converting them to RF signals through an additional Tx chain to cancel SI before LNA. Although digital-assisted A-SIC can effectively address the frequency selectivity of SI to combat multipath channels, it cannot eliminate the noise in the original Tx chain and will introduce additional noise, which is detrimental to D-SIC. Specifically, this implies that D-SIC will be ineffective in cancelling residual SI when Tx noise and additional noise predominate.

In contrast, the replica signal used by MTD A-SIC, derived after the PA of the Tx chain, contains both nonlinear components and Tx noise. This replica signal is split and directed to various taps with different delay, each equipped with a tunable attenuator and phase shifter. Ultimately, the signals from all the taps are superimposed in the Rx chain to cancel SI before LNA. Compared to the digital-assisted A-SIC, the unsubstitutable advantage of the MTD A-SIC is the ability to suppress Tx noise since the replica signal includes all distortions introduced in the Tx chain. Consequently, our theoretical analysis will focus on A-SIC based on the MTD architecture.

\subsection{Modeling of MTD A-SIC}
\begin{figure}[!t]
\centering
\includegraphics[width=3.0in]{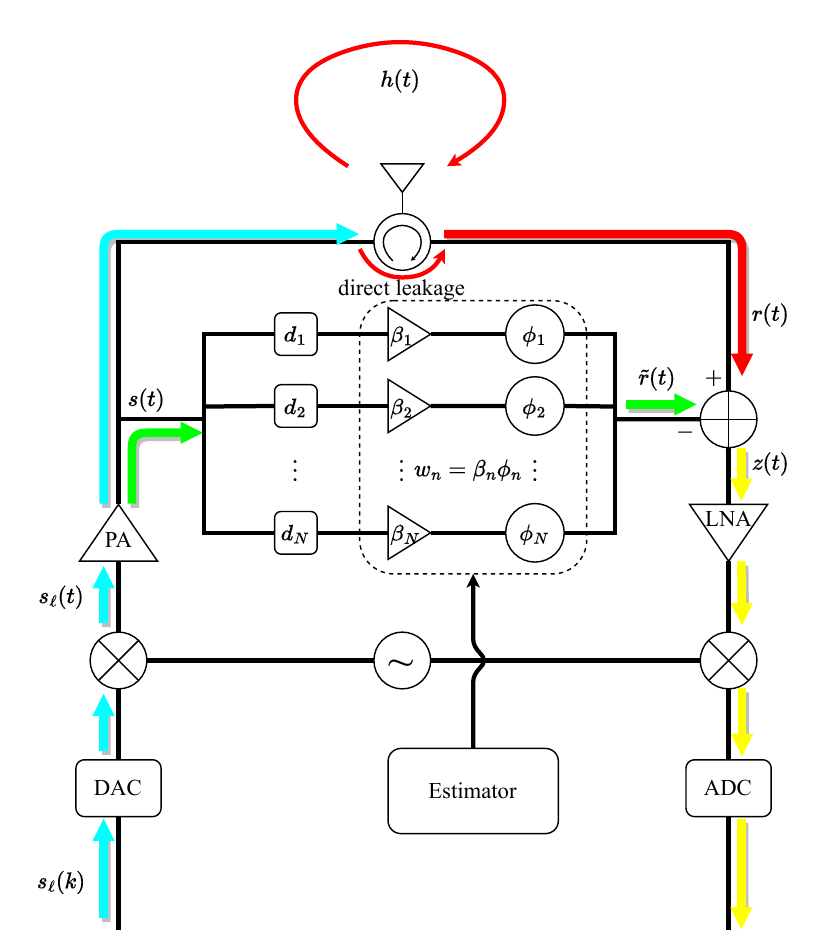}
\caption{ Analog self-interference cancellation based on MTD.}
\label{A-SIC_Fig}
\end{figure}
As illustrated in Fig. \ref{A-SIC_Fig}, the Tx signal $s(t)$ in RF can be expressed as
\begin{align}
s(t) &= \Re \left[ {s_\mathrm{b}(t){\mathrm{e}^{\jmath 2\pi {f_\mathrm{c}}t}}} \right]\ ,
\end{align}
where $f_\mathrm{c}$ is the carrier frequency and $s_\mathrm{b}(t)$ is the equivalent baseband signal of $s(t)$. Here, the impact of phase noise can be almost negligible in the case of an IBFD with a common LO for up- and down-conversion \cite{syrjala2014analysis}.

Taking into the distortion in Tx chain, such as PA nonlinearity, I/Q imbalance and Tx noise, $s_\mathrm{b}(t)$ can be decomposed as
\begin{align}
s_\mathrm{b}(t) = {{s_\mathrm{\ell}}(t) + {s_\mathrm{n\ell}}(t) + {n_\mathrm{t}}(t)}\ ,
\end{align}
i.e. the linear component ${s_\mathrm{\ell}}(t)$, the nonlinear component ${s_\mathrm{n\ell}}(t)$ and and the Tx noise ${n_\mathrm{t}}(t)$. 

Assuming that $s_\mathrm{b}(t)$ is of bandwidth $B$ and considering I/Q imbalance, ${s_\mathrm{\ell}}(t)$ can be expressed based on Shannon-Nyquist sampling theorem as
\begin{equation}\label{SINC}
s_\mathrm{\ell}(t) = \sum\limits_k {s_\mathrm{\ell,IQ}^*(k){\mathop{\rm sinc}\nolimits} (Bt - k)}\ ,    
\end{equation}
where the $k$th sampling point $s_\mathrm{\ell,IQ}(k)$ at Nyquist sampling rate can be expressed as
\begin{align}
    s_\mathrm{\ell,IQ}^*(k) = &s_\mathrm{\ell}(t)\delta(t-k/B) \notag\\
    = &b_0s_\mathrm{\ell}(k)+b_1s_\mathrm{\ell}^*(k)\ .
\end{align}
Here, ${s_\mathrm{\ell}}(k)$ is the $k$th Tx symbol and the image rejection ratio (IRR) of Tx chain can be expressed as $|b_{0}|^2/|b_{1}|^2$. Accordingly, the sampling points of ${s_\mathrm{n\ell}}(t)$ can be modeled as
\begin{align}
    s_\mathrm{n\ell}(k) =  &\sum_{\substack{p=3\\p = odd}
    }^{+\infty}c_{p}|s_\mathrm{\ell,IQ}(k)|^{p-1}s_\mathrm{\ell,IQ}(k) \notag\\
    =&\sum_{\substack{p=3\\p = odd}
    }^{+\infty}\sum_{q=0}^{p} {c_{p,q}s_\mathrm{\ell}^{q}(k)[s_\mathrm{\ell}^*(k)]^{p-q}}\ . 
\end{align}
where $c_{p}$ is the coefficient of $p$-order term of nonlinear components, and $c_{p,q}$ are determined by the coefficients of PA nonlinearity and I/Q imbalance. 

The Tx noise ${n_\mathrm{t}}(t)$ is modeled as a band-limited white Gaussian process, whose autocorrelation function of ${n_\mathrm{t}}(t)$ can be expressed as
\begin{align}
    \mathbb{E}(n_\mathrm{t}(t)n_\mathrm{t}^*(t+\tau)) = \rho_{n_\mathrm{t}}\mathrm{sinc}(B\tau)\ .
\end{align}

The Tx signal $s(t)$ experiences a multipath SI channel and is superposed by direct leakage of the circulator at the co-located receiver. Then, SI at the receiver can be described as
\begin{equation}
r(t) = {h(t)* s(t) + {{\alpha_0}s(t - {\tau _0})} }\ ,
\end{equation}
where ${h(t)}$ is the impulse response of the multipath SI channel, ${\alpha_0}$ and ${\tau_0}$ are the complex attenuation and the delay of direct leakage. Based on tapped delay line (TDL) model, ${h(t)}$ can be modeled as
\begin{equation}
{h(t)} = \Re \left[\sum\limits_{m = 1}^M{{\alpha_m}\delta(t - {\tau _m})}{\mathrm{e}^{\jmath2\pi {f_\mathrm{c}}t}}\right]\ , 
\end{equation}
where $M$ is the number of clusters in the SI channel and ${\tau_m}$ as well as ${\alpha_m}$ are the delay and the complex attenuation of paths in ${m}$th cluster respectively. Therefore, the baseband equivalent model of $r(t)$ can be expressed as
\begin{align}
r_\mathrm{b}(t) = &\sum\limits_{m = 0}^M{{\alpha_m}s_\mathrm{b}(t - {\tau _m})}\notag\\
= &h_\mathrm{SI}(t)* s_\mathrm{b}(t)\ ,
\end{align}
where the baseband equivalent model $h_\mathrm{SI}(t)$ of the SI channel can be expressed as
\begin{equation} 
h_\mathrm{SI}(t) = \sum\limits_{m = 0}^M{{\alpha_m}\delta(t - {\tau _m})}\ .   
\end{equation} 
Similarly, the baseband equivalent model of the reconstruction signal $\tilde r(t)$ generated by MTD A-SIC can be represented as
\begin{equation} \label{approx_SI}
\tilde r_\mathrm{b}(t) = \sum\limits_{n = 1}^N{{\beta_n}{\mathrm{e}^{\jmath{\phi_n}}}{\mathrm{e}^{-\jmath{2{\pi}f_\mathrm{c}d_n}}}s_\mathrm{b}(t - {d_n})}\ ,
\end{equation}
where $N$ is the number of taps, ${d_n}$ is the delay of the signal on ${n}$th tap, and ${\beta_n}$ as well as ${\phi_n}$ are the attenuation and the phase of the signal on ${n}$th tap respectively. For the sake of brevity, we denote the weight on the $n$th tap in A-SIC as $w_n = {\beta_n}{\mathrm{e}^{\jmath{\phi_n}}}$. Thus, we can rewritten $\tilde r_\mathrm{b}(t)$ as 
\begin{equation}
\tilde r_\mathrm{b}(t) = \tilde h_\mathrm{SI}(t)* s_\mathrm{b}(t)\ ,
\end{equation}
where $\tilde h_\mathrm{SI}(t)$ is the baseband equivalent impulse response of MTD A-SIC, which can be expressed as
\begin{equation}
\tilde h_\mathrm{SI}(t) = \sum\limits_{n = 1}^N{w_n{\mathrm{e}^{-\jmath{2{\pi}f_\mathrm{c}d_n}}}\delta(t - {d_n})}\ .
\end{equation}
Taking into account the estimation error of tap weights $w_n$, we define
\begin{equation}
\hat r_\mathrm{b}(t) = \sum\limits_{n = 1}^N{\hat w_n{\mathrm{e}^{-\jmath{2{\pi}f_\mathrm{c}d_n}}}s_\mathrm{b}(t - {d_n})}\ ,
\end{equation}
where $\hat{w}_n$ is the estimated value of $w_n$. Therefore, the baseband equivalent model of the residual SI $z(t)$ can be expressed as
\begin{align}\label{Residual_SI}
z_\mathrm{b}(t) = &  \underbrace{r_\mathrm{b}(t) + n_\mathrm{r}(t)}_{Received\ SI} - \underbrace{\hat r_\mathrm{b}(t)}_{Reconstructed\ SI} \notag \\ 
   = & \underbrace {r_\mathrm{b}(t)-\tilde r_\mathrm{b}(t)}_{Approximation\ error} + \underbrace {\tilde r_\mathrm{b}(t)-\hat r_\mathrm{b}(t)}_{Estimation\ error}+ \underbrace {n_\mathrm{r}(t)}_{Rx\ noise}\ ,
\end{align}
where ${n_\mathrm{r}}(t)$ is the Rx noise. As illustrated in Eq. \eqref{Residual_SI}, the error of SI based on MTD A-SIC can be divided into:
\begin{itemize}
    \item {\textit{Approximation error:}} The approximation error resulting from the reconstruction of SI by limited taps with non-ideality RF components is the inherent factor that influences the cancellation performance of MTD A-SIC. 
\end{itemize}
\begin{itemize}
    \item {\textit{Estimation error:}} The estimation error is influenced by the noise in the reference signal used by estimation algorithms such as the LMS algorithm.
\end{itemize}
\begin{itemize}
    \item {\textit{Rx noise:}} The Rx noise inherent in the Rx chain is usually 110 dB lower than the Tx signal in Wi-Fi systems.
\end{itemize}

In practice, the estimation error is generally significantly lower than the approximation error if there are no desired Rx signals in the reference signal used by the estimation algorithms, since the estimation error is primarily affected by Rx noise, which is 110 dB lower than SI \cite{huang2021alms,choi2013simultaneous,zhao2016impacts}. In terms of approximation error, the quantization bias resulting from the limited accuracy of the RF components, such as attenuators or ADCs in the feedback loop, can be reduced by improving the accuracy of the RF components \cite{choi2013simultaneous,zhao2016impacts,wang2016effect}. Therefore, in this paper, we only focus on the approximation error regardless of the estimation error and the quantization bias of tap weights, since it ultimately determines the limit of cancellation performance based on MTD A-SIC. 

\section{Theoretical Analysis\\ and Optimization}
In this section, we first analyze the approximation error for the cyclostationary SI based on MTD A-SIC in the deterministic multipath channel, which turns out to be equivalent to the approximation error of stationary Gaussian white process under the assumption of Gaussian symbols, greatly simplifying the performance analysis and optimization procedure. In addition, we provide the expression for the approximation error under the constraints on tap weights. Furthermore, from the perspective of stochastic multipath channels, we prove that the corresponding approximation error can be tightly bounded as the sum of the approximation error associated with each single-path. Next, we demonstrate that the optimization of the tap delays of MTD A-SIC with maximum tolerable approximation error can be seen as a collection of problems for minimizing the approximation error of MTD A-SIC under the constraints of different number of taps. Moreover, we propose a heuristic algorithm to estimate the optimal tap delays. Finally, we summarize the key factors that determine the fundamental limit of MTD A-SIC.


\subsection{Analysis of Approximation Error} \label{Appr. Error}
For the sake of notational brevity, we rewrite the approximated SI in Eq. \eqref{approx_SI} in the inner product form, i.e. 
\begin{equation}
  \tilde r_\mathrm{b}(t) = \boldsymbol{w}^\mathrm{T}\tilde{\boldsymbol{s}}(t),
\end{equation}
where $\boldsymbol{w}=[w_1\ \cdots \ w_N]^\mathrm{T}$, $\tilde{\boldsymbol{s}}(t) = \boldsymbol{D}\cdot[s_\mathrm{b}(t-d_1)\  \cdots \ s_\mathrm{b}(t-d_N)]^\mathrm{T}\  $, and $\boldsymbol{D}=\mathrm{diag}({\mathrm{e}^{-\jmath{2{\pi}f_\mathrm{c}d_1}}}\ \dots\ {\mathrm{e}^{-\jmath{2{\pi}f_\mathrm{c}d_N}}})$.
Consequently, the approximation error of SI can be rewritten as
\begin{align}
\varepsilon(t) = r_\mathrm{b}(t)-\boldsymbol{w}^\mathrm{T}\tilde{\boldsymbol{s}}(t)\ .
\end{align}
First, we consider the approximation error in a deterministic SI channel. 
\subsubsection{Deterministic SI channel}
\
\newline
\indent For a deterministic SI channel, the MMSE of $\varepsilon(t)$ can be expressed as
\begin{align}\label{MMSE_e}
\min\limits_{\boldsymbol{w}}\ \mathbb{E}\Big(\Big|r_\mathrm{b}(t)-\boldsymbol{w}^\mathrm{T}\tilde{\boldsymbol{s}}(t)\Big|^2\Big)\ .
\end{align}
\paragraph{No constraints on $\boldsymbol{w}$}
By taking the derivative of $\mathbb{E}(|\varepsilon(t)|^2)$, it can be easily derived that the optimal tap weights of Eq. \eqref{MMSE_e}, when $\boldsymbol{w} \in \mathbb{C}^N$, can be expressed as
\begin{align}
    \boldsymbol{w}_\mathrm{{mmse}} = ({\boldsymbol{R}}_{\tilde s}^\mathrm{T})^{-1}{\boldsymbol{R}}_{r\tilde s}^\mathrm{T}\ ,
\end{align}
where ${\boldsymbol{R}}_{r\tilde s} = \mathbb{E}(r_\mathrm{b}(t)\tilde{\boldsymbol{s}}^\mathrm{H}(t))$ is the cross-correlation matrix between $r_\mathrm{b}(t)$ and $\tilde{\boldsymbol{s}}(t)$, which can be expressed as
\begin{align}
    [\mathbb{E}(r_\mathrm{b}(t)s_\mathrm{b}^*(t-d_1))\  \cdots \ \mathbb{E}(r_\mathrm{b}(t)s_\mathrm{b}^*(t-d_N))]\cdot\boldsymbol{D}^\mathrm{H}\ .
\end{align} 
Similarly, ${\boldsymbol{R}}_{\tilde s}= \mathbb{E}(\tilde{\boldsymbol{s}}(t)\tilde{\boldsymbol{s}}^\mathrm{H}(t))$ is the autocorrelation matrix of $\tilde{\boldsymbol{s}}(t)$. 

Notably, $\boldsymbol{w}_\mathrm{{mmse}}$ is time-varying due to the cyclostationarity of ${s_\mathrm{\ell}}(t)$ and ${s_\mathrm{n\ell}}(t)$. Unfortunately, obtaining the real-time statistical characteristics of SI signals is not feasible in practical systems. As a compromise, according to Wiener-Khinchin theorem, we consider minimizing the time average of $\mathbb{E}(|\varepsilon(t)|^2)$, which can be expressed as 
\begin{align}\label{TA_MMSE_e}
    &\min\limits_{\boldsymbol{w}}\ \lim\limits_{T\rightarrow +\infty}\frac{1}{T}\int_{ - T}^{T}{\mathbb{E}\Big(|\varepsilon(t)|^2\Big)\mathrm{d}t}\notag\\
    =&\min\limits_{\boldsymbol{w}}\ \int_{ - \infty}^{+\infty} {{{\Big| {{H_\mathrm{SI}}(f) - {\tilde{H}_\mathrm{SI}(f)} } \Big|}^2P_s(f)}\mathrm{d}f}\ .
\end{align}
Here, ${H_\mathrm{SI}}(f)=\mathcal{F}[h_\mathrm{SI}(t)]$, ${\tilde{H}_\mathrm{SI}}(f)=\mathcal{F}[\tilde h_\mathrm{SI}(t)]$, and $P_s(f)$ is the power spectral density (PSD) of $s_\mathrm{b}(t)$ satisfying 
\begin{align}\label{rho_t}
   \rho_\mathrm{t}=\int_{ - \infty}^{+\infty} {P_s(f)}\mathrm{d}f\ , 
\end{align}
where $\rho_\mathrm{t}$ is the average Tx power. 

Particularly, the following theorem is provided.
\begin{thm}
    When ${{s_\ell}(k) \sim \mathcal{CN}(0,\rho_{\ell})}$ is uncorrelated with different sampling points and Tx noise $n_\mathrm{t}(t)$, the approximation error of SI (
    Eq. \eqref{TA_MMSE_e}) can be expressed as
    \begin{align}\label{TA_MMSE_eCN_f}
    &\min\limits_{\boldsymbol{w}}\ \lim\limits_{T\rightarrow +\infty}\frac{1}{T}\int_{ - T}^{T}{\mathbb{E}\Big(|\varepsilon(t)|^2\Big)\mathrm{d}t}\notag\\
    =&\min\limits_{\boldsymbol{w}}\ \frac{\rho_\mathrm{t}}{B}\int_{ - \frac{B}{2}}^{\frac{B}{2}} {{{\Big| {{H_\mathrm{SI}}(f) - {\tilde{H}_\mathrm{SI}(f)} } \Big|}^2}\mathrm{d}f}\ .     
    \end{align} 
    \begin{proof}
        See Appendix A.
    \end{proof}
\end{thm}
Especially, Eq. \eqref{TA_MMSE_eCN_f} can be regarded as a Wiener filtering process as follows:
\begin{align}\label{TA_MMSE_eCN_t}
    &\min\limits_{\boldsymbol{w}}\ \frac{\rho_\mathrm{t}}{B}\int_{ - \frac{B}{2}}^{\frac{B}{2}} {{{\Big| {{H_\mathrm{SI}}(f) - {\tilde{H}_\mathrm{SI}(f)} } \Big|}^2}\mathrm{d}f}\notag\\
    =&\min\limits_{\boldsymbol{w}}\ \rho_\mathrm{t}\mathbb{E}\Big(\Big|{(h_\mathrm{SI}(t)-\tilde h_\mathrm{SI}(t))*x(t)}\Big|^2\Big)\ ,
\end{align}
where $x(t)$ is a stationary Gaussian white process with autocorrelation function $R_x(\tau)=\mathrm{sinc}(B\tau)$. Similarly, We define $y(t)=h_\mathrm{SI}(t)*x(t)$ and
\begin{align}
    \tilde{\boldsymbol{x}}(t) = \boldsymbol{D}\cdot[x(t-d_1)\  \cdots \ x(t-d_N)]^\mathrm{T}\ .
\end{align}   
Then, the optimal tap weights of Eq. \eqref{TA_MMSE_eCN_t} can be derived as
\begin{align}\label{opt_w_u}
    \boldsymbol{w}^{\star}_\mathrm{ub} &= {\left({\boldsymbol{R}}_{\tilde x}^\mathrm{T}\right)}^{-1}{\boldsymbol{R}}_{y\tilde x}^\mathrm{T}\ ,
\end{align} 
where ${\boldsymbol{R}}_{\tilde{x}}=\mathbb{E}(\tilde{\boldsymbol{x}}(t)\tilde{\boldsymbol{x}}^\mathrm{H}(t))$ is the autocorrelation matrix of $\tilde{\boldsymbol{x}}(t)$, and ${\mathbf{R}}_{y\tilde{x}}=\mathbb{E}(y(t)\tilde{\boldsymbol{x}}^\mathrm{H}(t))$ is the cross-correlation matrix between $y(t)$ and $\tilde{\boldsymbol{x}}(t)$. 

Notably, $\mathbb{E}(|\varepsilon(t)|^2)$ can be divided into 
\begin{align}
    \mathbb{E}(|\varepsilon(t)|^2) = \mathbb{E}(|\varepsilon_{s_{\mathrm{\ell},\mathrm{n\ell}}}(t)|^2)+\mathbb{E}(|\varepsilon_{n_\mathrm{t}}(t)|^2)\ ,
\end{align}
where 
\begin{align}
\varepsilon_{s_{\mathrm{\ell},\mathrm{n\ell}}}(t) =& (h_\mathrm{SI}(t)-\tilde h_\mathrm{SI}(t))* (s_{\mathrm{\ell}}(t)+s_\mathrm{n\ell}(t)),\\
\varepsilon_{n_\mathrm{t}}(t) =& (h_\mathrm{SI}(t)-\tilde h_\mathrm{SI}(t))* n_\mathrm{t}(t) \ .
\end{align}
Interestingly, $\boldsymbol{w}^{\star}_\mathrm{ub}$ are also the optimal tap weights to minimum $\mathbb{E}(|\varepsilon_{n_\mathrm{t}}(t)|^2)$, which means that MTD A-SIC to minimum the approximation error in Eq. \eqref{TA_MMSE_e} is also the optimal eliminator of Tx noise $n_\mathrm{t}(t)$ under the assumption of Gaussian symbols.
\paragraph{Peak amplitude constraint on $\boldsymbol{w}$}
In practice, the tunable attenuators have limited power. Thus, taking into account the peak amplitude constraints on $\boldsymbol{w}$, Eq. \eqref{TA_MMSE_eCN_t} can be written as
\begin{align}\label{problem_w}
    \mathop {\min }\limits_{\boldsymbol{w}}\quad &{\mathbb{E}}\Big(\left|{y(t)-{\boldsymbol{w}}^\mathrm{T}\tilde{\boldsymbol{x}}(t)}\right|^2\Big)\\
    =\ &\mathbb{E}(|y(t)|^2)-\boldsymbol{w}^\mathrm{T}{\mathbf{R}}_{\tilde{x}y}-{\mathbf{R}}_{y\tilde{x}}\boldsymbol{w}^\mathrm{*}+\boldsymbol{w}^\mathrm{T}{\boldsymbol{R}}_{\tilde{x}}\boldsymbol{w}^\mathrm{*}\notag\\
    \quad \mathrm{s.t.}\quad  &{\left\| {\boldsymbol{w}} \right\|_\infty} \le w_0\notag\ .
\end{align}    
Eq. \eqref{problem_w} is a convex optimization problem and satisfies the Slate condition. Therefore, leveraging Lagrange multipliers, the optimal tap weights of Eq. \eqref{problem_w} can be derived as 
\begin{align}\label{opt_w}
    \boldsymbol{w}^{\star} &= {\left({\boldsymbol{R}}_{\tilde{x}}^\mathrm{T}+\boldsymbol{\Lambda}^{\star}\right)}^{-1}{\boldsymbol{R}}_{y\tilde{x}}^\mathrm{T}\ ,
\end{align} 
and the proof is provided in Appendix B. Here, $\boldsymbol{\Lambda}^{\star} = diag(\lambda_1^{\star}\ \cdots\ \lambda_N^{\star})$, where $\lambda_n^\star$ for $n=1,\cdots,N$ is the optimal Lagrange multiplier that satisfies the Karush-Kuhn-Tucker conditions in Eq. \eqref{KKT}. 

Therefore, let $\boldsymbol{w}=\boldsymbol{w}^\star$, the MMSE of the approximation error in Eq. \eqref{TA_MMSE_eCN_t} under power constraints on tap weights can be expressed as
\begin{align}\label{rho}
    \rho_\varepsilon=\rho_\mathrm{t}\Big(&\sum\limits_{i=0}^M\sum\limits_{j=0}^M\alpha_{i}\alpha_{j}^\mathrm{H}\mathrm{sinc}(B(\tau_{i}-\tau_{j}))-(\boldsymbol{w}^\star)^\mathrm{T}{\mathbf{R}}_{\tilde{x}y}\notag\\
    &-{\mathbf{R}}_{y\tilde{x}}(\boldsymbol{w}^\star)^\mathrm{*}+(\boldsymbol{w}^\star)^\mathrm{T}{\boldsymbol{R}}_{\tilde{x}}(\boldsymbol{w}^\star)^\mathrm{*}\Big)\ .
\end{align}

Since the SI channels are time-varying in nature, we will analyze the approximation error under stochastic SI channels next.
\subsubsection{Stochastic SI channel}
\
\newline
\indent Taking into account the uncertainty of the SI channel, without loss of generality, we assume that ${\alpha_m \sim \mathcal{CN}(0,a_m^2)}$ for $m=1,\dots,M$ is uncorrelated with that at different $m$, where $a_m^2$ is the average path attenuation of the paths in $m$th cluster. Especially, the expectations for the constant term $\alpha_0$ are $\mathbb{E}(\alpha_0)=a_0$ and $\mathbb{E}(\alpha_0^2)=a_0^2$. Unfortunately, it is hard to calculate the expectation of $\rho_{\varepsilon}$ for $\alpha$, which is denoted as $\bar\rho_{\varepsilon}$, since we cannot obtain the close-form expression of $\boldsymbol{\Lambda}^\star$. 
Nevertheless, a special case arises when $\boldsymbol{w} \in \mathbb{C}^N$, which leads to a lower bound of $\bar\rho_{\varepsilon}$.
\paragraph{The lower bound of $\bar\rho_{\varepsilon}$} 
Especially, Eq. \eqref{opt_w_u} can be rewritten as
\begin{align}
    \boldsymbol{w}_\mathrm{ub}^\star=&{\left({\boldsymbol{R}}_{\tilde{x}}^\mathrm{T}\right)}^{-1}\sum_{m=0}^M\alpha_m\mathbb{E}(x(t-\tau_m)\tilde{\boldsymbol{x}}^\mathrm{H}(t))^\mathrm{T}\notag\\
    =&\sum_{m=0}^M\alpha_m{\left({\boldsymbol{R}}_{\tilde{x}}^\mathrm{T}\right)}^{-1}{{\boldsymbol{R}}}_{x_{m}\tilde{x}}^\mathrm{T}\ ,
\end{align}
where ${{\boldsymbol{R}}}_{x_{m}\tilde{x}}=\mathbb{E}(x(t-\tau_m)\tilde{\boldsymbol{x}}^\mathrm{H}(t))$. We denote 
\begin{align}
    \tilde{\boldsymbol{w}}^\star_{\mathrm{ub},m} = {\left({\boldsymbol{R}}_{\tilde{x}}^\mathrm{T}\right)}^{-1}{{\boldsymbol{R}}}_{x_{m}\tilde{x}}^\mathrm{T}\ , 
\end{align}
and then $\rho_{\varepsilon}$ without constraint on $\boldsymbol{w}$ can be expressed as
\begin{align}
    \rho_{\varepsilon} = \rho_\mathrm{t}\mathbb{E}\Big(\Big|\sum_{m=0}^M\alpha_m(x(t-\tau_m)-(\tilde{\boldsymbol{w}}^\star_{\mathrm{ub},m})^\mathrm{T}\tilde{\boldsymbol{x}}(t))\Big|^2\Big)\ .
\end{align}
Since $\tilde{\boldsymbol{w}}^\star_{\mathrm{ub},m}$ are unrelated with $\alpha_m(m=0,\dots,M)$, the expectation of $\rho_{\varepsilon}$ for $\alpha$ can be expressed as
\begin{align}
    \bar\rho_{\varepsilon} = \rho_\mathrm{t}\sum_{m=0}^Ma_m^2\mathbb{E}\Big(\Big|(x(t-\tau_m)-(\tilde{\boldsymbol{w}}^\star_{\mathrm{ub},m})^\mathrm{T}\tilde{\boldsymbol{x}}(t))\Big|^2\Big)\ .
\end{align}
We define
\begin{align}\label{e_m}
    \varepsilon_{m}(t) &= x(t-\tau_m)-\boldsymbol{w}^\mathrm{T}\tilde{\boldsymbol{x}}(t)\ ,    
\end{align}
and the MMSE of $\varepsilon_{m}(t)$ when $\boldsymbol{w} \in \mathbb{C}^N$ can be derived as
\begin{align}\label{E_e_m_lb}
    {\bar \varepsilon}^2_{m,\mathrm{lb}} = 1-2\Re\Big[(\tilde{\boldsymbol{w}}^\star_{\mathrm{ub},m})^\mathrm{T}{\boldsymbol{R}}_{\tilde{x}x_{m}}\Big]+\Big\|{\boldsymbol{R}}_{\tilde{x}}^{\frac{1}{2}}(\tilde{\boldsymbol{w}}^\star_{\mathrm{ub},m})^\mathrm{*}\Big\|_2^2\ ,
\end{align}
where $\tilde{\boldsymbol{w}}^\star_{\mathrm{ub},m}$ are the optimal tap weights. Therefore, the lower bound of $\bar\rho_{\varepsilon}$ with  unconstrained  $\boldsymbol{w}$ can be expressed as
\begin{align}\label{lower bound}
    \bar\rho_{\varepsilon} \ge \rho_\mathrm{t}\sum_{m=0}^Ma_m^2{\bar \varepsilon}^2_{m,\mathrm{lb}}\ ,
\end{align}
where the equality holds when $\boldsymbol{w}^\star=\boldsymbol{w}^\star_\mathrm{ub}$. 

This lower bound has the good property that the MMSE of the approximation error in statistical multipath channels can be characterized by the sum of that in each single-path. Similarly, we hope that the upper bound also has the same property.
\paragraph{The upper bound of $\bar\rho_{\varepsilon}$} Considering the constraints $\|\boldsymbol{w}\|_\infty\le w_{\varepsilon_{m}}$ in Eq. \eqref{e_m}, the optimal tap weights can be derived as
\begin{align}\label{opt_w_m_lb}
    \tilde{\boldsymbol{w}}^{\star}_{\mathrm{lb},m} &= {\left({\boldsymbol{R}}_{\tilde{x}}^\mathrm{T}+\boldsymbol{\Lambda}^{\star}_m\right)}^{-1}{\boldsymbol{R}}_{x_{m}\tilde{x}}^\mathrm{T}\ .
\end{align} 
and we define
\begin{align}\label{opt_w_lb}
    {\boldsymbol{w}}^{\star}_{\mathrm{lb}}=\sum_{m=0}^M\alpha_m\tilde{\boldsymbol{w}}^{\star}_{\mathrm{lb},m}\ .
\end{align}
According to the norm inequality, when $w_{\varepsilon_{m}} \le \frac{w_0}{(M+1)|\alpha_m|}$, the power constraints $\|{\boldsymbol{w}}^{\star}_{\mathrm{lb}}\|_\infty \le w_0$ will hold. To make $\tilde{\boldsymbol{w}}^{\star}_{\mathrm{lb},m}$ unrelated with $\alpha_m$, let $w_{\varepsilon_{m}}=\frac{w_0}{(M+1)\sqrt{a_m^2}}$,
and the probability that $\|{\boldsymbol{w}}^{\star}_{\mathrm{lb}}\|_\infty\le w_0$ can be derived as
\begin{align}\label{P_w_lb}
    P\{\|{\boldsymbol{w}}^{\star}_{\mathrm{lb}}\|_\infty\le w_0\}\ge1-2Q(\frac{M+1}{\sqrt{M}}) \ ,
\end{align}
where $Q(\cdot)$ is the complementary cumulative distribution function (CDF) of the standard normal distribution and the proof is provided in Appendix C. The right-hand side of Eq. \eqref{P_w_lb} is monotonically increasing with $M$, which is greater than 0.95 as $M=1$.
We denote 
\begin{align}
    \beta_M = 1-2Q(\frac{M+1}{\sqrt{M}})\ ,
\end{align}
and define the MMSE of $\varepsilon_{m}(t)$ when $\|\boldsymbol{w}\|_\infty\le \frac{w_0}{(M+1)\sqrt{a_m^2}}$ as
\begin{align}\label{E_e_m_ub}
    {\bar \varepsilon}^2_{m,\mathrm{ub}} = 1-2\Re\Big[(\tilde{\boldsymbol{w}}^\star_{\mathrm{lb},m})^\mathrm{T}{\boldsymbol{R}}_{\tilde{x}x_{m}}\Big]+\Big\|{\boldsymbol{R}}_{\tilde{x}}^{\frac{1}{2}}(\tilde{\boldsymbol{w}}^\star_{\mathrm{lb},m})^\mathrm{*}\Big\|_2^2\ ,
\end{align}
Then, the upper bound of $\bar\rho_{\varepsilon}$ can be obtained by
\begin{align}\label{upper_bound}
    \rho_\mathrm{t}\sum_{m=0}^Ma_m^2{\bar \varepsilon}^2_{m,\mathrm{ub}}\ge \beta_M\bar\rho_{\varepsilon} \ .
\end{align}
Therefore, the lower and upper bound of $\bar\rho_{\varepsilon}$ can be expressed as
\begin{equation}\label{bound}
     \rho_\mathrm{t}\sum_{m=0}^Ma_m^2{\bar \varepsilon}^2_{m,\mathrm{lb}} \le \bar\rho_{\varepsilon} \le \frac{\rho_\mathrm{t}}{\beta_M}\sum_{m=0}^Ma_m^2{\bar \varepsilon}^2_{m,\mathrm{ub}}\ ,
\end{equation}
In this way, the analysis of the approximation error in statistical multipath channels can be simplified to the analysis of the approximation error in single-path.

Notably, since Tx noise can only be suppressed in A-SIC, the upper bound of the overall SIC ability can be expressed as
\begin{equation}
    {\frac{\rho_\mathrm{t}}{\rho_{n_\mathrm{t}}\sum\limits_{m=0}^Ma_m^2{\bar \varepsilon}^2_{m,\mathrm{lb}}}}\ .
\end{equation}

\subsection{Optimization of MTD A-SIC}
In this subsection, we provide  the optimization procedures for obtaining the tap delays based on the previous theoretical analysis. 

\begin{figure}[!t]
\centering
\includegraphics[width=3.2in]{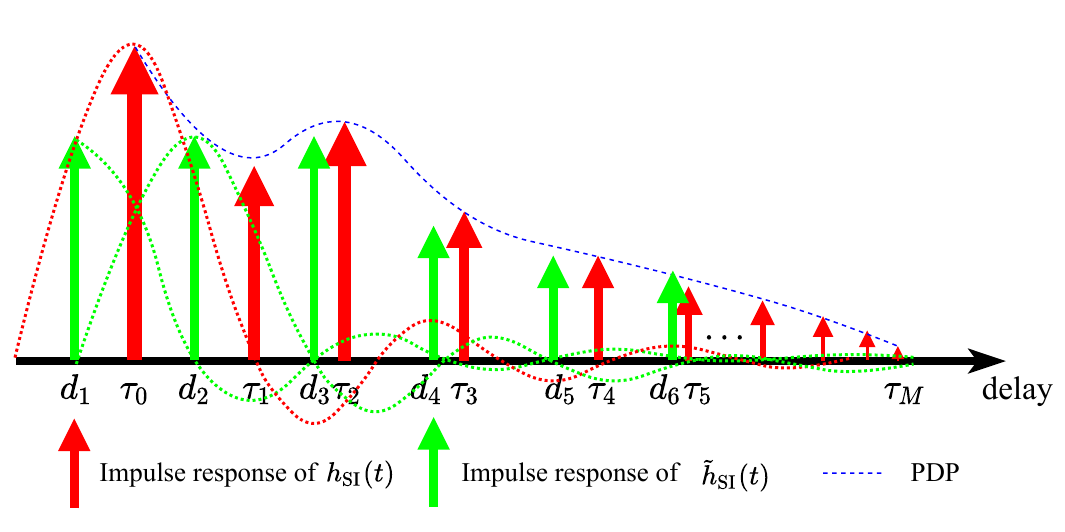}
\caption{Reconstruction of SI channel response in time domain.}
\label{Reconstruction}
\end{figure}

As illustrated in Fig. \ref{Reconstruction}, from the perspective of sampling theorem, MTD A-SIC reconstructs a bandwidth-limited signal at time $t-\tau_m$ by the sampling points of the signal at time $t-d_n$ for $n=1,\cdots,N$.
Defining the tap delays of MTD A-SIC as $\boldsymbol{d}=[d_1 \ \cdots \ d_N]^{\mathrm{T}}$, it is intuitive that the optimal tap delays are $\boldsymbol{d}^\star=[\tau_1 \ \cdots \ \tau_M]^\mathrm{T}$ when $N=M$, where the approximation error will be 0 when the tap weights are $\boldsymbol{w}=[\alpha_1 \ \cdots \ \alpha_M]^{\mathrm{T}}$. 

However, in practical systems the delay profile of SI channels normally change with time, while the tap delays $\boldsymbol{d}$ need to be kept constant for the sake of stability and simplicity. This means a MTD A-SIC which is robust to channel variations is highly desirable. Moreover, from the perspective of implementation cost, we hope to minimize the total number of taps in the MTD A-SIC on the condition of ensuring the SIC performance.

Considering the variation of the SI channels and assuming that $\tau_m \in \mathcal{T}$, the upper bound of $\bar\rho_{\varepsilon}$ can be further expressed as 
\begin{align}\label{up_bound}
    \bar\rho_{\varepsilon}\le &\frac{\rho_\mathrm{t}}{\beta_M}\sum_{m=0}^Ma_m^2{\bar \varepsilon}^2_{m,\mathrm{ub}}\notag\\
    \le &\frac{(M+1)\rho_\mathrm{t}}{\beta_M}\max_{\tau_m \in \mathcal{T}}a_m^2{\bar \varepsilon}^2_{m,\mathrm{ub}}\ .
\end{align}
Then, the problem on the design for MTD A-SIC with minimum number of taps, under the constraint of maximum allowable approximation error with power $\varepsilon_0^2$, can be described as
\begin{align}\label{MINLP}
    \mathop{\min_{N\in \mathbb{Z}^+ }}  \ \ &N \  \\
     \mathrm{s.t.}\ \  &\mathop{\max_{\tau_m \in \mathcal{T}}} \ a_m^2{\bar \varepsilon}^2_{m,\mathrm{ub}}(\boldsymbol{d}) \le \frac{\beta_M\varepsilon_0^2}{(M+1)\rho_\mathrm{t} } \ ,\notag\\
     &\boldsymbol{d}\in \mathbb{R}_+^N\ .\notag\\
     \ 
\end{align}
For notational brevity, we denote that
\begin{align}\label{beta}
    \eta = \frac{\beta_M\varepsilon_0^2}{(M+1)\rho_\mathrm{t}}\ .
\end{align}
 Eq. \eqref{MINLP} is a standard mixed-integer nonlinear programming (MINLP) problem, for which the Karush-Kuhn-Tucker conditions of the optimal solution ($N^\star \ne 0$) can be expressed as
\begin{align} \label{KKT_N}
\begin{cases}
    &\mathop{\min\limits_{\boldsymbol{d}\in \mathbb{R}_+^{N^\star}}}\quad\ \mathop{\max\limits_{\tau_m \in \mathcal{T}}} \ a_m^2{\bar \varepsilon}^2_{m,\mathrm{ub}}(\boldsymbol{d}) \le \eta\ ,\\
    &\mathop{\min\limits_{\boldsymbol{d}\in \mathbb{R}_+^{(N^{\star}-1)}}}\mathop{\max\limits_{\tau_m \in \mathcal{T}}} \ a_m^2{\bar \varepsilon}^2_{m,\mathrm{ub}}(\boldsymbol{d}) > \eta\ .
\end{cases}    
\end{align}
It is hard to solve the KKT condition (Eq. \eqref{KKT_N}) directly, since we cannot provide the closed-form expression for ${\bar \varepsilon}^2_{m,\mathrm{ub}}(\boldsymbol{d})$. Instead, we provide a iterative method (on the number of taps $N$) when an initial value of the tap delays, i.e. $\boldsymbol{d}_0$ is available. We summarize the corresponding procedures in \textbf{Algorithm} 1.
\begin{algorithm}
    \caption{Finding the Optimal Tap Delays  $\boldsymbol{d}^\star$}
    \label{alg:1}
    \begin{algorithmic}    
        \STATE \textbf{Input:\ } $\boldsymbol{d}_0$ \text{\quad (CALL \textbf{Algorithm} 2)}
        \STATE $N=\mathrm{dim}(\boldsymbol{d}_0)$
        \STATE $\boldsymbol{d}_N^\star=\mathop{\arg\min}\limits_{\boldsymbol{d}\in \mathbb{R}_+^{N}}\mathop{\max\limits_{\tau_m \in \mathcal{T}}} \ a_m^2{\bar \varepsilon}^2_{m,\mathrm{ub}}(\boldsymbol{d})$ 
        \WHILE{\ $\mathop{\max\limits_{\tau_m \in \mathcal{T}}} \ a_m^2{\bar \varepsilon}^2_{m,\mathrm{ub}}(\boldsymbol{d}_N^\star)\le \eta$\ } 
        \STATE $\boldsymbol{d}^\star=\boldsymbol{d}_N^\star$
        \STATE $N = N-1$;
        \STATE $\boldsymbol{d}_N^\star=\mathop{\arg\min}\limits_{\boldsymbol{d}\in \mathbb{R}_+^{N}}\mathop{\max\limits_{\tau_m \in \mathcal{T}}} \ a_m^2{\bar \varepsilon}^2_{m,\mathrm{ub}}(\boldsymbol{d})$ 
        \ENDWHILE
        \RETURN $\boldsymbol{d}^\star$
    \end{algorithmic}
\end{algorithm}

Notably, the approximation error can be divided into: {\textit{path attenuation $a_m^2$}} and {\textit{interpolation error ${\bar \varepsilon}^2_{m,\mathrm{ub}}(\boldsymbol{d})$}}. Furthermore, the constraint in Eq. \eqref{MINLP} can be rewritten as
\begin{align}\label{Cons_e_m_ub}
    \mathop{\max\limits_{\tau_m \in \mathcal{T}}} \ {\bar \varepsilon}^2_{m,\mathrm{ub}}(\boldsymbol{d}) \le \frac{\eta}{a_m^2}\ ,
\end{align}
which reveals that MTD A-SIC has different requirements for interpolation error with different path attenuations. Especially, the requirement of the  SIC performance associated  with a path whose attenuation is weaker than $\eta$ can be as low as 0. 

\begin{figure*}[!t]
\centering
\subfloat[]{\includegraphics[width=3in]{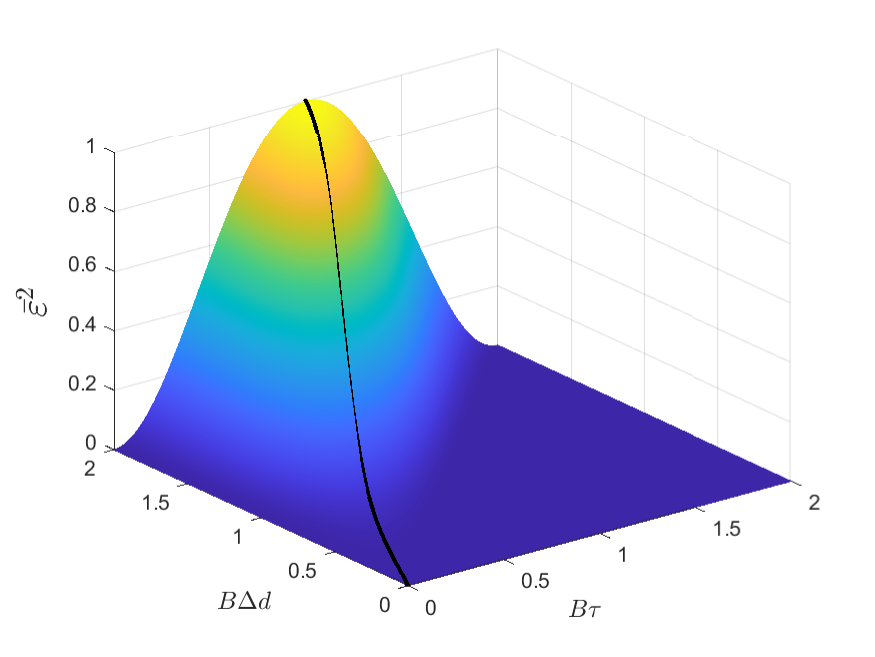}
\label{max_error_1}}
\hfil
\subfloat[]{\includegraphics[width=3in]{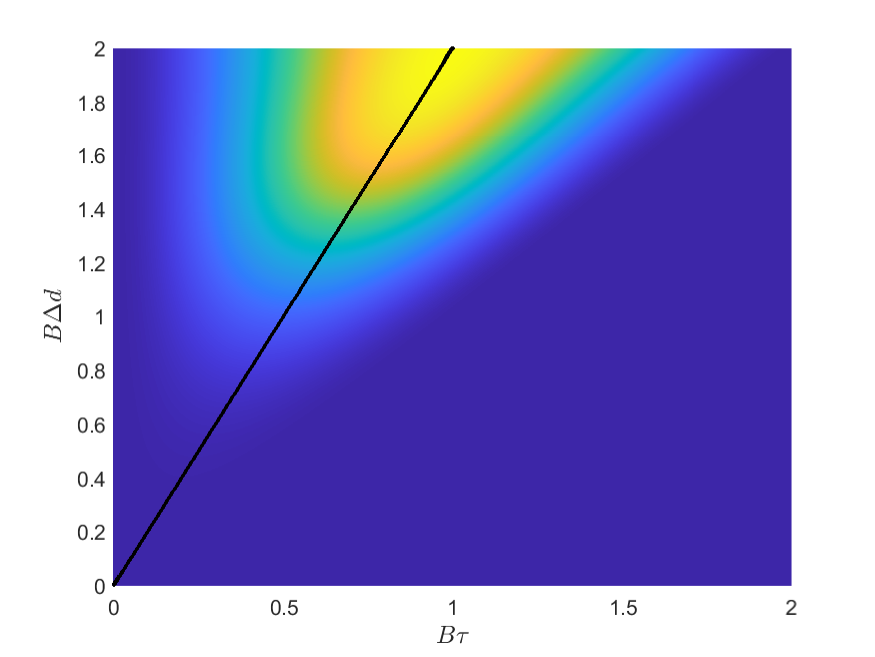}
\label{max_error_2}}
\caption{The graph of ${\bar \varepsilon}^2_{m,\mathrm{ub}}(\boldsymbol{d}_2)$ when $\boldsymbol{d}_2=[0,\Delta d]^\mathrm{T}$ and $\tau \in [0,\Delta d]$, where the black line is the maximum value curve. (a) 3-D graph. (b) 2-D graph.}
\label{max_error}
\end{figure*}
\begin{figure}[!t]
\centering
\includegraphics[width=3.3in]{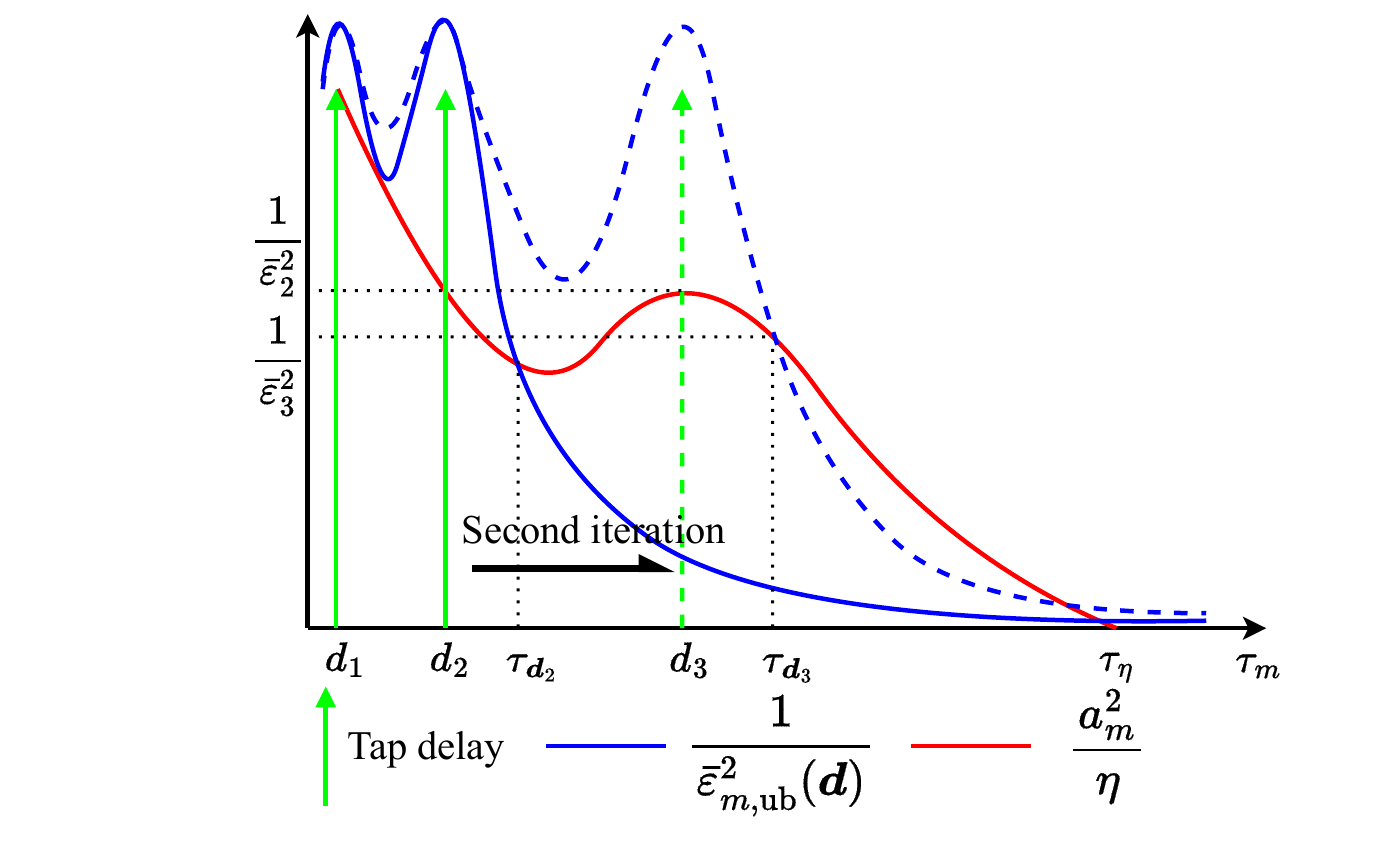}
\caption{Diagram of the second iteration in Algorithm 2.}
\label{Fig_Alg_2}
\end{figure}

To demonstrate the influence of different requirements on the interpolation error, we take MTD A-SIC with tap delays $\boldsymbol{d}_2=[d_1\ d_2]^\mathrm{T}$ as an example. As proved in Appendix D, when the constraint in Eq. \eqref{opt_w_m_lb} satisfies  
\begin{align}\label{Con_w_m}
    w_{\varepsilon_{m}}=\frac{w_0}{(M+1)\sqrt{a_m^2}}\ge1.27\ ,
\end{align}  
${\bar \varepsilon}^2_{m,\mathrm{ub}}(\boldsymbol{d}_2)$ will be equal to ${\bar \varepsilon}^2_{m,\mathrm{lb}}(\boldsymbol{d}_2)$, thus  we can provide the close-form expression for ${\bar \varepsilon}^2_{m,\mathrm{ub}}(\boldsymbol{d}_2)$. Notably, Eq. \eqref{Con_w_m} holds when the path attenuation $a_m^2\le$ -28.6 dB, $w_0$ = 1 and $M$ = 20, which is common in most multipath environments. Based on the assumption of Eq. \eqref{Con_w_m}, as illustrated in Fig. \ref{max_error}, when $B(d_2-d_1)=B \Delta d \in (0,2] $ and $\tau_m \in[d_1,d_2]$, the maximum of ${\bar \varepsilon}^2_{m,\mathrm{ub}}(\boldsymbol{d})$,  can be expressed as
\begin{equation}\label{max_error_m}
    \mathop{\max}_{\tau_m \in[d_1,d_2]}\ {\bar \varepsilon}^2_{m,\mathrm{ub}}(\boldsymbol{d}_2) = 1-\frac{2\mathrm{sinc}^2(B\Delta d/2)}{1+\mathrm{sinc}(B\Delta d)}\ ,
\end{equation}
where $\tau_m=\frac{d_1+d_2}{2}$. Notably, Eq. \eqref{max_error_m}  increases monotonically with the increase of $\Delta d$, implying that the tap delays will exhibit non-uniformity due to the distinct path attenuation. 
Defining that 
\begin{align}
    \tau_{\eta}&= \inf \{\tau\ |\ \mathop{\max}_{\tau_m \ge \tau} a_m^2\le \eta,\tau \in \mathcal{T}\}\ ,\\
    \tau_{\boldsymbol{d}} &= \inf \{\tau_m\ |\ a_m^2{\bar \varepsilon}^2_{m,\mathrm{ub}}(\boldsymbol{d}) \ge \eta,\tau_m \in \mathcal{T}\}\ ,
\end{align}
and leveraging the non-uniformity of tap delays, we develop Algorithm \ref{alg:2} to determine the initial tap delays $\boldsymbol{d}_0$, where the diagram of the algorithm is illustrated in Fig. \ref{Fig_Alg_2}.
\begin{algorithm}
    \caption{Determining the Initial Tap Delays $\boldsymbol{d}_0$}
    \label{alg:2}
    \begin{algorithmic}    
        \IF{$\tau_{\eta} \ne \inf \mathcal{T}$ }
        \STATE $N=1$
        \STATE $\boldsymbol{d}_N=d_1$
        \WHILE{\ $\tau_{\boldsymbol{d}_N} \neq \varnothing$\ }
        \STATE $N=N+1$
        \STATE $\bar\varepsilon^2_N = \mathop {\min}\limits_{\tau_{\boldsymbol{d}_N}\le\tau_m\le\tau_{\eta}}\  \frac{\eta}{a_m^2}$
        \STATE $\Delta d_N = \sup \{\Delta d\ |\ 1-\frac{2\mathrm{sinc}^2(B\Delta d/2)}{1+\mathrm{sinc}(B\Delta d)} \le \bar\varepsilon^2_N\} $ 
        \STATE $d_N = d_{N-1}+\Delta d_N$
        \STATE $\boldsymbol{d}_N=[\boldsymbol{d}_{N-1}^\mathrm{T}\ d_N]^\mathrm{T}$
        \ENDWHILE
        \RETURN $\boldsymbol{d}_0=\boldsymbol{d}_N$
        \ELSE 
        \RETURN $N^\star=0$
        \ENDIF
    \end{algorithmic}
\end{algorithm}

\subsection{Key Factors for MTD A-SIC Performance}
Through the above theoretical analysis, we are able to identify the key factors that have significant impact on the performance of MTD A-SIC in multipath SI channels.

\subsubsection{Tap delays} Tap delays $\boldsymbol{d}$ are of critical importance for the performance of MTD A-SIC  because the MTD architecture for A-SIC is in essence a reconstruction of the multipath channel of SI. Thus in the most favorable scenarios where the tap delays coincide with the path delays of SI channel, the reconstruction error can be as low as zero.

\subsubsection{PDP of SI channels} For the same reason as the above, the PDP of the multipath SI channel also plays a critical role in the reconstruction  (or approximation) error. 
Therefore,it is very beneficial for employing the spatial beamforming or reconfigurable intelligence surfaces (RIS) \cite{tewes2022full} in order to modify the PDP of the effective SI channel to our advantage.

\subsubsection{Carrier frequency} According to ${\bar \varepsilon}^2_{m,\mathrm{lb}}$, the carrier frequency does not affect the interpolation error of SI when there are no constraints on the tap weights. However, it is noteworthy that higher frequencies result in greater attenuation of the radio waves in the free space. This suggests that in scenarios such as millimeter-wave communication \cite{singh2020millimeter}, SI will experience greater path attenuation, which is beneficial for reducing the approximation error.
\subsubsection{Bandwidth} On the positive side, when bandwidth doubles, the Rx noise floor will increase 3 dB, implying that the performance requirements for the overall SIC can be relaxed as the Tx power remains constant. However, the thermal noise in the transmitter will also increase 3 dB accordingly, which means that the minimum performance requirement for A-SIC will remain unchanged overall. Unfortunately, according to Eq. \eqref{max_error_m}, it indicates that the interpolation error based on MTD A-SIC will increase with the bandwidth as the delay interval $\Delta d$ remains constant.
\subsubsection{Passive A-SIC} Although passive A-SIC cannot eliminate multipath components in SI, it can suppress direct leakage, which typically corresponds to the strongest path in the SI. Therefore, effective passive A-SIC can help to compensate for insufficient cancellation performance of MTD A-SIC with regard to the direct leakage, thereby relaxing the demands on the cancellation performance of MTD A-SIC, and in turn reduce the complexity of A-SIC.

 \section{Simulations and Discussions}
\begin{table}[!t]
\caption{configurations of Wi-Fi IBFD radio \label{IBFD radio}}
\centering
\begin{tabular}{c|c}
\hline
Parameter & Value\\
\hline
Protocol stack &  802.11 ax\\
\hline
Bandwidth & 80 MHz\\
\hline
MCS & 64 QAM 3/4\\
\hline
Tx power & 20 dBm\\
\hline
Power of the nonlinear components & -10 dBm\\
\hline
Carrier frequency &  5.6 GHz\\
\hline
Tx IRR & 25 dB\\
\hline
Tx SNR & 60 dB\\
\hline
Circulator attenuation  & -25 dB\\
\hline
Delay of direct leakage  & 0.4 ns\\
\hline
Rx noise floor & -90 dBm\\
\hline
\end{tabular}
\end{table}
In this section, we provide simulation results on the validation for  our theoretical results on MTD A-SIC as well as the performance of our proposed optimization algorithms. In addition, we present some discussion on the implications of our  theoretical results and optimization algorithms. 

\subsection{Performance Metric and Simulation Setup}

The main metric to evaluate the performance of SIC is the signal cancellation ratio ($\mathrm{SCR}$), which is defined as follows:
\begin{equation}\label{SCR}
    \mathrm{SCR} =\frac{\rho_\mathrm{t}}{\bar\rho_{\varepsilon},}\ 
\end{equation}
where $\rho_\mathrm{t}, \rho_{\varepsilon}$ denotes the power of the Tx signal, and the residual error after SIC, respectively.

\begin{figure}[!t]
\centering
\includegraphics[width=2.8in]{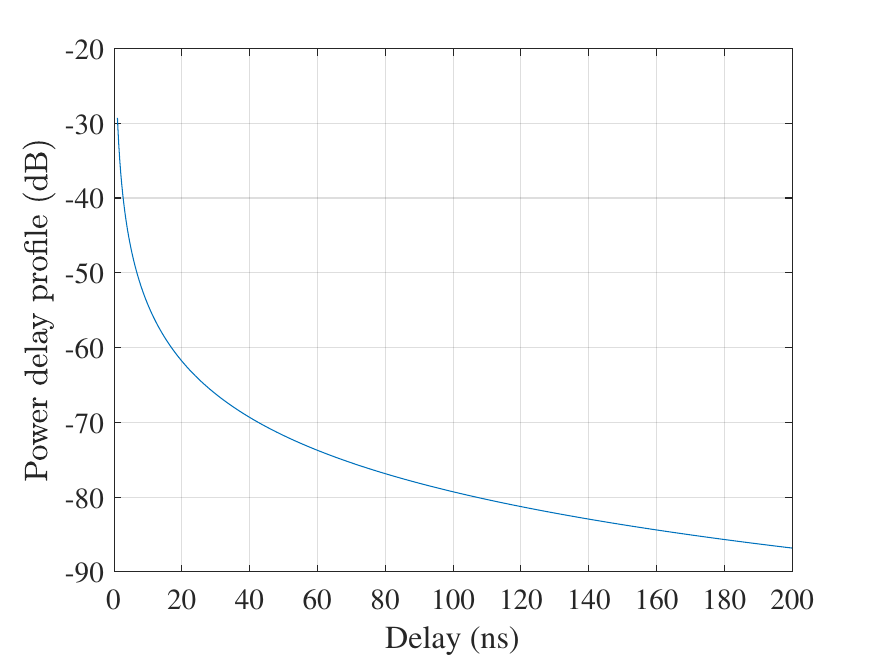}
\caption{Power-delay profile.}
\label{PDP}
\end{figure}

In all our simulations, we employ Wi-Fi IBFD radios whose configurations are as listed in TABLE \ref{IBFD radio}. Moreover, three types of TDL channel model (TDL-A, TDL-B and TDL-C) in 3GPP \cite{ETSI3GPP} are used for modeling the SI channel, whose normalized delays (to the delay spread $\tau_{\mathrm{DS}}$) are listed in TABLE \ref{TDL-model}. Furthermore, as demonstrated in Fig. \ref{PDP}, we employ the PDP to model the path attenuation as follows \cite{wu2014power}:
\begin{equation}\label{PDP_Model}
    a_m^2(\tau_m)\ (\mathrm{[dB]}) = -254.29-25\log_{10}(\tau_m)\ .
\end{equation}

\begin{table}[!t]
\caption{Normalized delays of TDL model \label{TDL-model}}
\centering
\begin{tabular}{c|c|c|c}
\hline
Tap & \makecell[c]{Normalized delays \\ TDL-A} & \makecell[c]{Normalized delays \\ TDL-B} & \makecell[c]{Normalized delays \\ TDL-C}\\
\hline
1 & 0.3819 & 0.1072 & 0.2099\\
\hline
2 & 0.4025 & 0.2155 & 0.2219\\
\hline
3 & 0.5868 & 0.2095 & 0.2329\\
\hline
4 & 0.4610 & 0.2870 & 0.2176\\
\hline
5 & 0.5375 & 0.2986 & 0.6366\\
\hline
6 & 0.6708 & 0.3752 & 0.6448\\
\hline
7 & 0.5750 & 0.5055 & 0.6560\\
\hline
8 & 0.7618 & 0.3681 & 0.6584\\
\hline
9 & 1.5375 & 0.3697 & 0.7935\\
\hline
10 & 1.8978 & 0.5700 & 0.8213\\
\hline
11 & 2.2242 & 0.5283 & 0.9336\\
\hline
12 & 2.1718 & 1.1021 & 1.2285\\
\hline
13 & 2.4942 & 1.2756 & 1.3083\\
\hline
14 & 2.5119 & 1.5474 & 2.1704\\
\hline
15 & 3.0582 & 1.7842 & 2.7105\\
\hline
16 & 4.0810 & 2.0169 & 4.2589\\
\hline
17 & 4.4579 & 2.8294 & 4.6003\\
\hline
18 & 4.5695 & 3.0219 & 5.4902\\
\hline
19 & 4.7966 & 3.6187 & 5.6077\\
\hline
20 & 5.0066 & 4.1067 & 6.3065\\
\hline
21 & 5.3043 & 4.2790 & 6.6374\\
\hline
22 & 9.6586 & 4.7834 & 7.0427\\
\hline
23 & 10.0000 & 5.0000 & 8.6523\\
\hline
\end{tabular}
\end{table}

\subsection{Effectiveness of Theoretical A-SIC Performance}

To validate the effectiveness of our theoretical results on the MTD A-SIC performance, we first provide the \textit{theoretical} SIC performance according to our analytic results in Sec. \ref{Appr. Error} as well as the \textit{numerical} SIC performance when the tap delays of MTD A-SIC are specified by \cite{bharadia2013full}, whose tap delays are listed in the TABLE \ref{A-SIC}.

\begin{table}[!t]
\caption{parameters of MTD A-SIC \label{A-SIC}}
\centering
\begin{tabular}{c|c}
\hline
Parameter & Value\\
\hline
Number $N$ of taps  & 8\\
\hline
Delay interval $\Delta d$ between adjacent taps  & 0.1 ns\\
\hline
Minimum delay of taps & 0.2 ns\\
\hline
Maximum value of attenuator & 0 dB\\
\hline
\end{tabular}
\end{table}

In our simulations, we first generate a Tx signal based on the configurations in TABLE \ref{IBFD radio}. Then, the SI signal is generated by the Tx signal experiencing the SI channel, whose actual delay is determined by a given delay spread $\tau_{\mathrm{DS}}$. Meanwhile, an MTD A-SIC with tap delays in TABLE \ref{A-SIC} processes the replicas of the Tx signal to generate the reconstructed SI signal, where the tap weights are calculated by the CVX toolbox based on the SI signal without Rx noise. Finally, the SI signal reconstructed by the MTD A-SIC is subtracted from the real SI signal to obtain the residual SI. 

\begin{figure}[!t]
\centering
\subfloat[]{\includegraphics[width=2.8in]{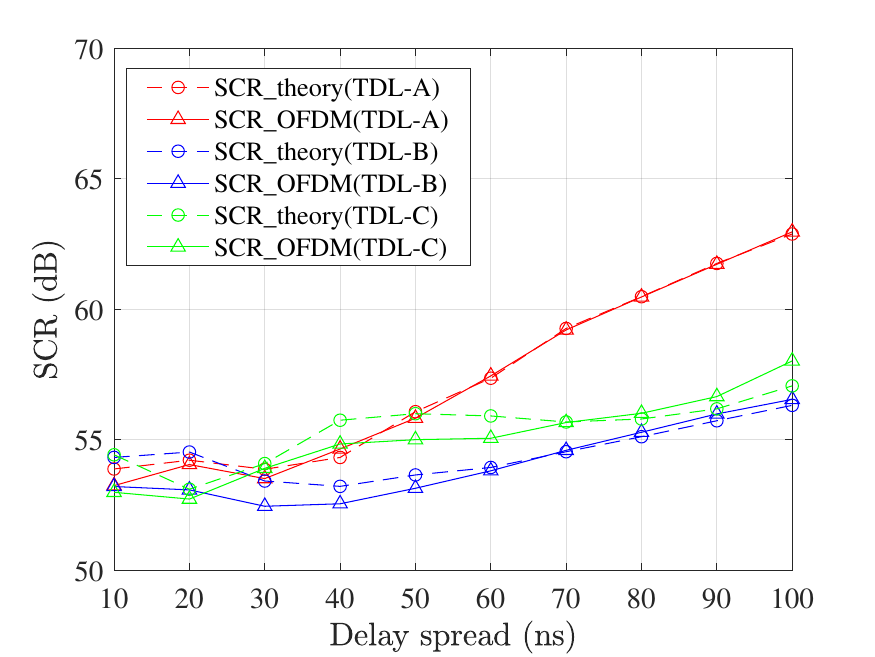}}
\hfil
\subfloat[]{\includegraphics[width=2.8in]{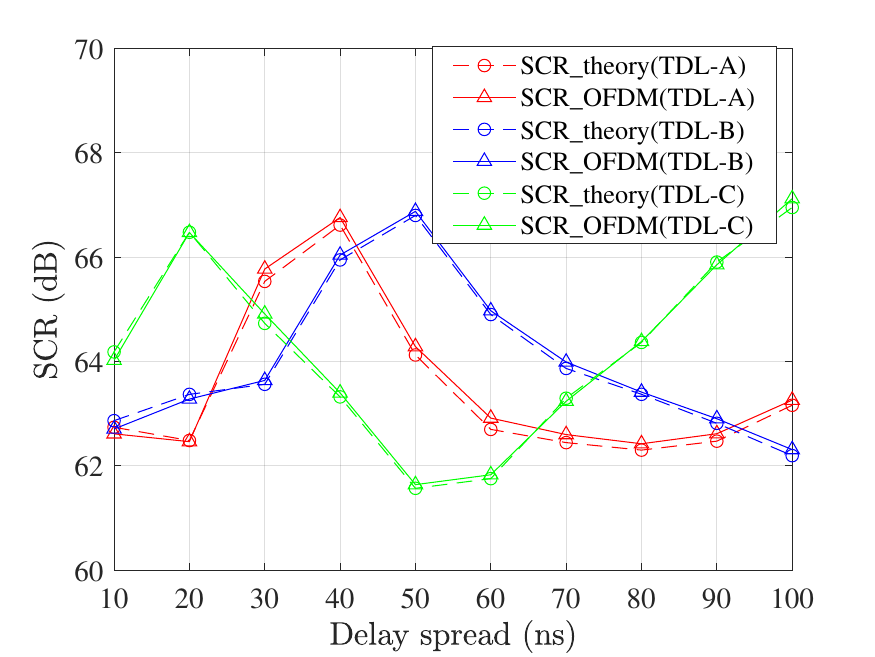}}
\hfil
\caption{Comparisons of $\mathrm{SCR}$ based on MTD A-SIC with different tap delays across various multipath SI channels. (a) The tap delays $\boldsymbol{d}_\mathrm{u}$ of MTD in \cite{bharadia2013full}, whose parameters are listed in TABLE \ref{A-SIC}. (b) The tap delays $\boldsymbol{d}_0$ of MTD optimized through Algorithm \ref{alg:2}.}
\label{SCR_ASIC}
\end{figure}

Comparisons of $\mathrm{SCR}$ corresponding to the numerical results via simulation and the theoretical results  are presented in Fig. \ref{SCR_ASIC}(a). It can be seen from Fig. \ref{SCR_ASIC}(a) that the SIC performance predicted by our theoretical result matches quite well with the actual performance. Moreover, it is worth noting that the SIC ability can achieve no less than 52 dB across various multipath channels, higher than the SIC performance reported in \cite{bharadia2013full}, which is approximately only 45 dB. The main reason for the performance improvement lies in the incorporation of phase shifters in our architecture, which is absent, however, in \cite{bharadia2013full}.



\subsection{Performance of the Optimized MTD A-SIC}
Next, we will focus on demonstrating the performance of 
our proposed optimization algorithms for the tap delays of MTD.

At first, we need to determine the parameters in Algorithm \ref{alg:2}. To achieve a total SIC ability of 110 dB, where the maximum cancellation performance of D-SIC is assumed to be 56 dB, we have to ensure that MTD A-SIC could contribute at least 54 dB. Therefore, according to Eq. \eqref{beta}, we can obtain $\eta=-67.6$ dB. Besides, considering the delay jitter of the direct leakage and multipath components, we set $d_1=0.2$ ns and $\mathcal{T}=\{\tau\ |\ \tau \ge 1\ \mathrm{ns}\}$ to ensure the robustness of the optimized MTD A-SIC to various SI channels.

The tap delays of MTD A-SIC optimized by Algorithm \ref{alg:2} are $\boldsymbol{d}_0$ = [0.2 0.6099 2.6624 9.7061 22.2061]$^\mathrm{T}$ ns. And the performance of the resulting  MTD A-SIC is illustrated in Fig. \ref{SCR_ASIC}(b). As demonstrated in Fig. \ref{SCR_ASIC} (b), $\mathrm{SCR}$ based on the optimized MTD A-SIC can achieve no less than 61.6 dB and has less fluctuations across various SI channels. Compared to the MTD A-SIC with tap delays in TABLE \ref{A-SIC}, this represents a 16\% increase in the worst-case performance and a 37.5\% reduction of the number of taps. 
\subsubsection{SIC Performance vs. the Delay Spread of the SI Channel}
\begin{figure}[!t]
\centering
\includegraphics[width=2.8in]{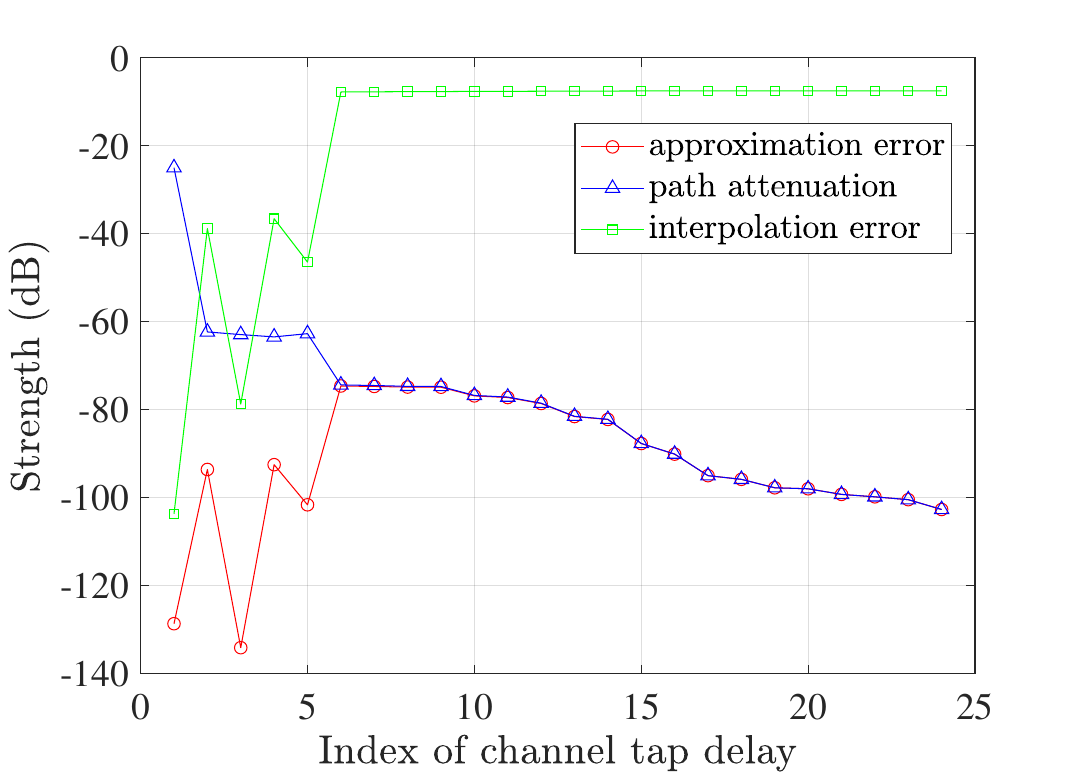}
\caption{The strength of the approximation error, the interpolation error, and the path attenuation associated with each path for MTD A-SIC.}
\label{Q1-2}
\end{figure}

Notably, $\mathrm{SCR}$ fluctuates across various SI channels with different $\tau_\mathrm{DS}$ due to the fact that there are two conflicting forces which together determine the level of the approximation error: one is \textit{interpolation error} associated with each path, which tends to \textit{increase} with larger delay; the other is the \textit{path attenuation} of each path, which \textit{decrease} with delay. Since the approximation error for each path is the product of its corresponding interpolation error and path attenuation, and the overall approximation error for MTD A-SIC is the sum of the approximation error for each path, the final $\mathrm{SCR}$ thus fluctuates with the $\tau_\mathrm{DS}$. Taking TDL-C channel with $\tau_\mathrm{DS}=100$ ns as an example, the strength of the approximation error, the interpolation error, and the path attenuation based on MTD A-SIC with tap delay $\boldsymbol{d}_0$ associated with each path are illustrated in Fig. \ref{Q1-2}.  

\subsubsection{SIC Performance vs. the Bandwidth}

\begin{figure}[!t]
\centering
\subfloat[]{\includegraphics[width=2.8in]{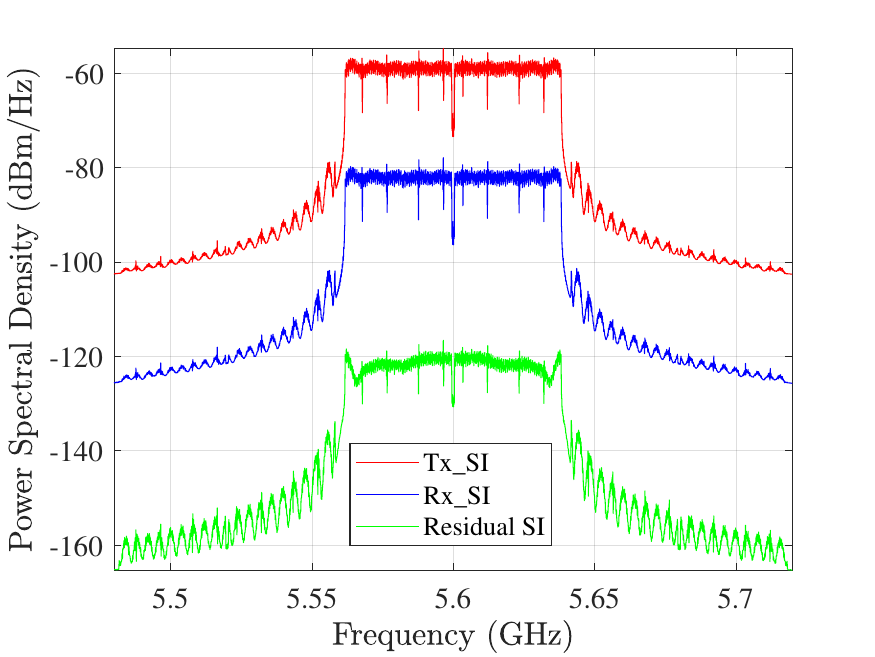}}
\label{PSD_80MHz}
\hfil
\subfloat[]{\includegraphics[width=2.8in]{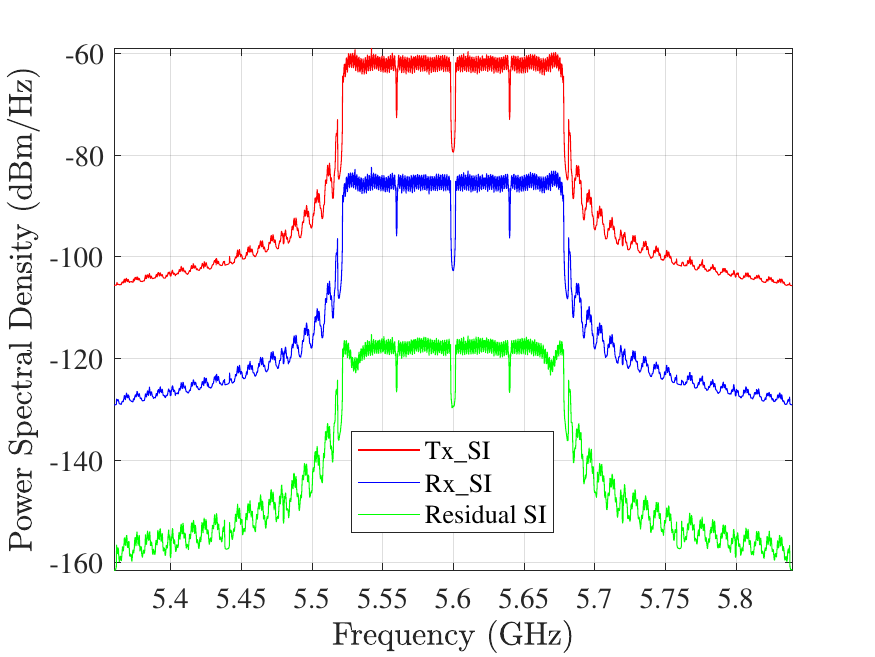}}
\label{PSD_160MHz}
\hfil
\caption{The PSD at different stages of SI with different bandwidth in channel B with $\tau_{\mathrm{DS}}=10$ ns when $\boldsymbol{d}=\boldsymbol{d}_0$. (a) 80 MHz. (b) 160 MHz.}
\label{PSD_80+160}
\end{figure}

Next, we assess the impact of the SI signal bandwidth on the approximation error based on the MTD A-SIC with $\boldsymbol{d}_0$. PSDs at different stages of SI, under TDL-B channel with $\tau_\mathrm{DS}=10$ ns, are presented in Fig. \ref{PSD_80+160}. The results indicate that as the bandwidth increases from 80 MHz to 160 MHz, the cancellation performance of the MTD A-SIC decreases from 62 to 55 dB. Obviously, the bandwidth has a significant impact on the performance of MTD A-SIC since the interpolation error of SI will decrease as bandwidth increases due to the constant tap delays. Therefore, the challenges of deploying IBFD will be more significant in the future with higher bandwidths.

\subsubsection{The Optimality of Our Heuristic Algorithm}
\begin{figure}[!t]
\centering
\includegraphics[width=2.8in]{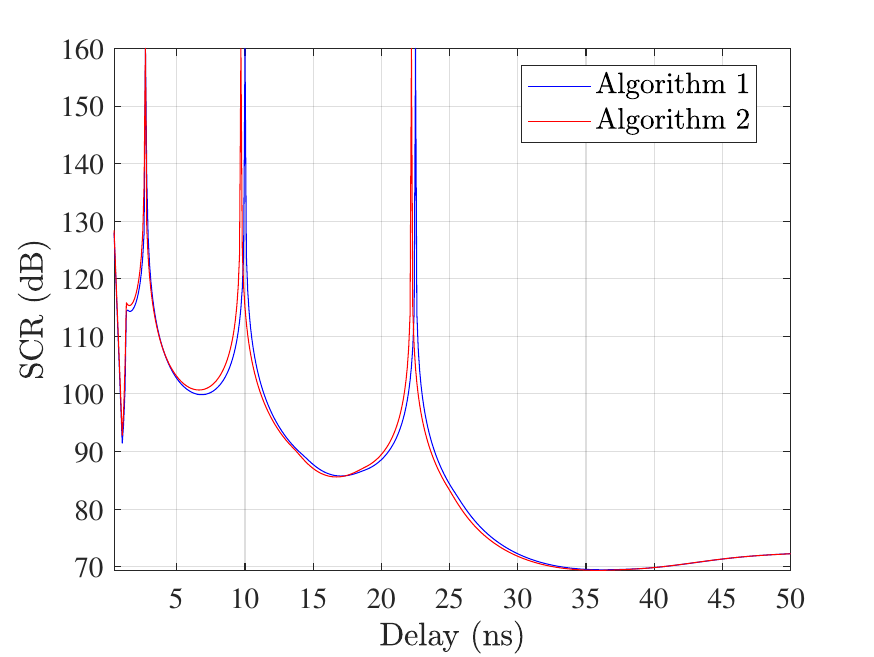}
\caption{The comparison of $1/{a_m^2{\bar \varepsilon}_{m,\mathrm{ub}}^2(\boldsymbol{d})}$ when $\boldsymbol{d}=\boldsymbol{d}_0$ and $\boldsymbol{d}=\hat{\boldsymbol{d}}^\star$.}
\label{SCR_Alg}
\end{figure}

In order to evaluate the optimality of our proposed heuristic algorithm (i.e. \textbf{Algorithm} 2) for obtaining the initial tap delays, we conduct a performance comparison between it and the MINLP algorithm in MATLAB in Fig. \ref{SCR_Alg}. It can be seen that the  tap delay $\boldsymbol{d}^\star$ output by the MATLAB algorithm is very close to the output of our heuristic algorithm $\boldsymbol{d}_0$, both in values of the tap delays and the $\mathrm{SCR}$ performance. This implies that our heuristic algorithm can achieve nearly the same performance as the complicated algorithm for general MINLP problems with far less complexity, and more importantly, our heuristic algorithm can provide structural insights on the optimal tap delays for the MDT A-SIC.


\subsection{Discussions}
\textit{Can SI be completely eliminated in the analog domain?} Theoretically, SI can be completely eliminated if the tap delays are perfectly aligned with those of the multipath components in SI. However, for SI in varying multipath channels, even achieving cancellation performace of no less than 80 dB is challenging due to limited coverage of tap delays. For example, as illustrated in Fig. \ref{PDP}, achieving the performance of 80 dB requires the coverage of tap delays exceeding 100 ns. Notably, with each doubling of the channel delay, path attenuation in Eq. \eqref{PDP_Model} will decrease by approximately 7.5 dB. This implies that a requirement of 110 dB of SIC would necessitate the coverage of tap delays exceeding 1600 ns, which is unrealistic. Therefore, we recommend that the design of MTD A-SIC should strike a balance between complexity and performance based on the specific context.

\section{Conclusion}
In this paper, we explore the performance limit of MTD A-SIC by taking into account practical factors such as cyclostationarity of the modulated signal, nonlinear distortions of the Tx chain, general multipath (rather than single-path) SI channels, and peak amplitude of the tap weights. By proving that the approximation error for the cyclostationary Tx signal is equivalent to that of a stationary Gaussian white process, the performance analysis involved can be greatly simplified. Moreover, we demonstrate that the approximation error under the general stochastic multipath channel can be tightly bounded by the sum of approximation error associated with each single-path. By capitalizing on the above structural results, a heuristic and efficient algorithm is provided to obtain the optimal MTD, especially the optimal tap delays, under the constraint of maximum approximation error. We believe that the theoretical analysis and optimization algorithms presented in this paper would provide valuable guidelines for the design of practical IBFD radios.


\begin{appendices}
\section*{Appendix A}
Defining $\mathrm{rect}(x)$ as
\begin{align}
    \mathrm{rect}(x)=
    \begin{cases}
        1 \quad |x|\le 1 \ ,\\
        0 \quad |x| > 1\ ,
    \end{cases}
\end{align}
and then $P_s(f)$, when ${s_\ell}(k)$ is uncorrelated with transmitter noise $n_\mathrm{t}(t)$, can be divided as
\begin{align}
    P_s(f) = P_{s_{\ell,\mathrm{n}\ell}}(f)+P_n(f)\ ,
\end{align}
where the PSD of $n_\mathrm{t}(t)$ can be expressed as
\begin{align}
    P_n(f) = \frac{\rho_{n_\mathrm{t}}}{B}\mathrm{rect}(\frac{2f}{B})\ ,
\end{align}
and the PSD of ${s_\ell}(t)+{s_{\mathrm{n}\ell}}(t)$, denoted as $P_{s_{\ell,\mathrm{n}\ell}}(f)$, can be expressed as
\begin{align}\label{S(f)}
    &\mathop{\lim}_{T\rightarrow +\infty} {\frac{\mathbb{E}\left({\Big| {\mathcal{F}\Big[({s_\ell}(t)+{s_{\mathrm{n}\ell}}(t))\mathrm{rect}(\frac{2t}{T})\Big]} \Big|}^2\right)}{T}}\notag\\
    =&\lim_{T\rightarrow +\infty}\frac{1}{T}\sum\limits_{k_1}\sum\limits_{k_2}\bigg( \mathcal{F}\left[\mathrm{sinc}(Bt-k_1)\mathrm{rect}(\frac{2t}{T})\right]\ \cdot \notag\\
    & \mathcal{F}^*\left[\mathrm{sinc}(Bt-k_2)\mathrm{rect}(\frac{2t}{T})\right]\ \cdot\sum_{\substack{p_1=1\\p_1 = odd}
    }^{+\infty}\sum_{\substack{p_2=1\\p_2 = odd}
    }^{+\infty}c_{p_1}c_{p_2}^* \notag\\
    & \mathbb{E}\left( {|s_\mathrm{\ell,IQ}(k_1)|^{p_1-1}s_\mathrm{\ell,IQ}(k_1)} 
     {|s_\mathrm{\ell,IQ}(k_2)|^{p_2-1}s_\mathrm{\ell,IQ}^*(k_2)}\right)\bigg)\notag\\
    =&\mathrm{rect}(\frac{2f}{B})\lim_{T\rightarrow +\infty}\frac{1}{TB^2}\sum\limits_{k_1}\sum\limits_{k_2}{\mathrm{e}^{\jmath 2\pi\frac{k_2-k_1}{B}f}}\sum_{p_1}\sum_{p_2}c_{p_1}c_{p_2}^*\notag\\
    & \mathbb{E}\left( {|s_\mathrm{\ell,IQ}(k_1)|^{p_1-1}s_\mathrm{\ell,IQ}(k_1)} 
     {|s_\mathrm{\ell,IQ}(k_2)|^{p_2-1}s_\mathrm{\ell,IQ}^*(k_2)}\right)\ .
\end{align}
Since ${{s_\ell}(k) \sim \mathcal{CN}(0,\rho_{\ell})}$ is uncorrelated with different sampling points, it can be proved that ${{s_{\ell,\mathrm{IQ}}}(k)}$ is also a independent circularly symmetric Gaussian variable, which satisfy ${{{s_{\ell,\mathrm{IQ}}}(k)} \sim \mathcal{CN}(0,(b_0^2+b_1^2)\rho_{\ell})}$.
Thus, there has
\begin{align}
    \mathbb{E}\left( {|s_\mathrm{\ell,IQ}(k)|^{p-1}s_\mathrm{\ell,IQ}(k)} 
     \right) &= \mathbb{E}\left( {[s_\mathrm{\ell,IQ}^*(k)]^{\frac{p-1}{2}}[s_\mathrm{\ell,IQ}(k)]^{\frac{p+1}{2}}} 
     \right) \notag\\ 
     &= 0\ . 
\end{align}
Furthermore, the expectation in Eq. \eqref{S(f)} can be expressed as
\begin{align}
    &\mathbb{E}\left( {|s_\mathrm{\ell,IQ}(k_1)|^{p_1-1}s_\mathrm{\ell,IQ}(k_1)} 
     {|s_\mathrm{\ell,IQ}(k_2)|^{p_2-1}s_\mathrm{\ell,IQ}^*(k_2)}\right) \notag\\ 
     =& \frac{\sigma^2_{p_1,p_2}}{c_{p_1}c_{p_2}^*} \delta(k_1-k_2)\ ,
\end{align}
where
\begin{align}
    \sigma^2_{p_1,p_2} =& c_{p_1}c_{p_2}^*\mathbb{E}\left( {|s_\mathrm{\ell,IQ}(k)|^{p_1+p_2}}\right) \notag\\
    =&c_{p_1}c_{p_2}^*(\frac{p_1+p_2}{2})![(b_0^2+b_1^2)\rho_{\ell}]^{\frac{p_1+p_2}{2}}\ .
\end{align}
Let $T=\frac{i}{B}$, where $i \sim \mathbb{N}_+$, according to the constraint of $P_s(f)$ in Eq. \eqref{rho_t}, $P_s(f)$ can be rewritten as
\begin{align}
   P_s(f)= &\mathrm{rect}(\frac{2f}{B})\bigg(\mathop{\lim}_{{i} \rightarrow +\infty}\frac{1}{iB}\sum_{k=1}^i \sum_{p_1}\sum_{p_2}\sigma_{p_1,p_2}^2+\frac{\rho_{n_\mathrm{t}}}{B}\bigg) \notag\\
   =&\frac{\rho_\mathrm{t}}{B}\mathrm{rect}(\frac{2f}{B})\ ,
\end{align}
which remains constant within $[-\frac{B}{2},+\frac{B}{2}]$. Therefore, Eq. \eqref{TA_MMSE_e} can be expressed as
    \begin{align}\label{P_AprError_xf_apx}
    \min\limits_{\boldsymbol{w}}\ \frac{\rho_\mathrm{t}}{B}\int_{ - \frac{B}{2}}^{\frac{B}{2}} {{{\Big| {{H_\mathrm{SI}}(f) - {\tilde{H}_\mathrm{SI}(f)} } \Big|}^2}\mathrm{d}f}\ .     
    \end{align} 

\section*{Appendix B}
According to Eq. \eqref{problem_w}, the problem of finding $\boldsymbol{w}^\star$ can be rewritten as
\begin{align}\label{problem_w_apx}
    \ \ \mathop {\arg\ \mathrm{min} }\limits_{\boldsymbol{w}}\quad &\boldsymbol{w}^\mathrm{T}{\boldsymbol{R}}_{\tilde{x}}\boldsymbol{w}^\mathrm{*}-\boldsymbol{w}^\mathrm{T}{\boldsymbol{R}}_{\tilde{x}y}-{\boldsymbol{R}}_{y\tilde{x}}\boldsymbol{w}^\mathrm{*}\\
    \quad\quad \mathrm{s.t.}\quad\quad\  &{\left\| {\boldsymbol{w}} \right\|_\infty} \le w_0\notag\ .
\end{align}
Here the autocorrelation matrix ${\boldsymbol{R}}_{{\tilde{x}}}=\mathbb{E}(\tilde{\boldsymbol{x}}(t)\tilde{\boldsymbol{x}}^\mathrm{H}(t))$ can be expressed as
\begin{align}
    \boldsymbol{D}\cdot
    \begin{bmatrix}
        1 & \cdots & \mathrm{sinc}(B(d_1-d_N))\\
        \mathrm{sinc}(B(d_2-d_1)) & \cdots & \mathrm{sinc}(B(d_2-d_N))\\
        \vdots & \ddots & \vdots\\
        \mathrm{sinc}(B(d_N-d_1)) & \cdots & 1
    \end{bmatrix}\cdot \boldsymbol{D}^\mathrm{H}\ ,
\end{align}
and the cross-correlation matrix ${\boldsymbol{R}}_{y{\tilde{x}}}=\mathbb{E}(y(t)\tilde{\boldsymbol{x}}^\mathrm{H}(t))$ can be expressed as
\begin{align}
    =&\sum_{m = 0}^M
    \begin{bmatrix}
        \alpha_m\mathrm{sinc}(B(\tau_m-d_1))\\
        \alpha_m\mathrm{sinc}(B(\tau_m-d_2))\\
        \vdots\\
        \alpha_m\mathrm{sinc}(B(\tau_m-d_N))
    \end{bmatrix}\cdot\mathbf{D}^\mathrm{H}\ .
\end{align}
Obviously, Eq. \eqref{problem_w_apx} is a convex optimization problem with strong duality, since it satisfies the Slater condition and ${\boldsymbol{R}}_{{\boldsymbol{xx}}}\succ 0$. Furthermore, the Karush-Kuhn-Tucker conditions can be expressed as
\begin{align}\label{KKT}
\begin{cases}
    &\lambda_n^\star(w_0^2-(w_n^\star)^2) = 0,\quad n = 1,\dots,N,\\
    &\|\boldsymbol{w}^\star\| \le w_0,\quad\quad\quad\quad\ \ n = 1,\dots,N,\\
    &\lambda_n^\star \ge 0,\quad\quad\quad\quad\quad\quad\ \ n = 1,\dots,N,\\
    &\boldsymbol{w}^{\star} = {\left({\boldsymbol{R}}_{\tilde{x}}^\mathrm{T}+\boldsymbol{\Lambda}^{\star}\right)}^{-1}{\boldsymbol{R}}_{y\tilde{x}}^\mathrm{T},
\end{cases}        
\end{align}
and the optimal solution of Eq. \eqref{problem_w_apx} can be expressed as 
\begin{align}
    \boldsymbol{w}^{\star} = {\left({\boldsymbol{R}}_{\tilde{x}}^\mathrm{T}+\boldsymbol{\Lambda}^{\star}\right)}^{-1}{\boldsymbol{R}}_{y\tilde{x}}^\mathrm{T}\ .
\end{align} 

\section*{Appendix C}
For the tap weights in Eq. \eqref{opt_w_lb}, since $\alpha_0 = a_0$ and $\alpha_m \sim \mathcal{CN}(0,a_m^2)$ for $m=1,\dots,M$, the distribution of the $n(n=1,\dots,N)$th element of ${\boldsymbol{w}}^{\star}_{\mathrm{lb}}=\sum_{m=0}^M\alpha_m\tilde{\boldsymbol{w}}^{\star}_{\mathrm{lb},m}$ satisfies 
\begin{align}
    w^{\star}_{\mathrm{lb},n}\sim \mathcal{CN}(a_0\tilde{w}^{\star}_{\mathrm{lb},0,n},\sum\limits_{m=1}^Ma_m^2|\tilde{w}^{\star}_{\mathrm{lb},m,n}|^2)\ ,
\end{align}
where $\tilde{w}^{\star}_{\mathrm{lb},m,n}$ is the $n$th element of $\tilde{\boldsymbol{w}}^{\star}_{\mathrm{lb},m}$ in Eq. \eqref{opt_w_m_lb}. Then the probability about $w^{\star}_{\mathrm{lb},n}$ can be derived as
\begin{align}
    &P\{|w^{\star}_{\mathrm{lb},n}-a_0\tilde{w}^{\star}_{\mathrm{lb},0,n}|\le w_0\}\notag\\
    =&1-2Q(\frac{w_0}{\sqrt{\sum\limits_{m=1}^Ma_m^2(\tilde{w}^{\star}_{\mathrm{lb},m,n})^2}})\ ,
\end{align}
Since $|\tilde{w}^{\star}_{\mathrm{lb},m,n}|\le w_{\varepsilon_{m}}$, when $w_{\varepsilon_{m}}=\frac{w_0}{(M+1)\sqrt{a_m^2}}$, there has
\begin{align}
    P\{\|{\boldsymbol{w}}^{\star}_{\mathrm{lb}}\|_\infty\le w_0\}&\ge P\{\max\limits_n|w^{\star}_{\mathrm{lb},n}|\le \frac{Mw_0}{M+1}\}\notag\\
    &\ge P\{\max\limits_n|w^{\star}_{\mathrm{lb},n}-a_0\tilde{w}^{\star}_{\mathrm{lb},0,n}|\le w_0\}\notag\\
    &\ge1-2Q(\frac{M+1}{\sqrt{M}}) \ .
\end{align}
\section*{Appendix D}
Defining $f_m=\mathrm{sinc}(B(\tau_m-d_1))$ and $g_m=\mathrm{sinc}(B(\tau_m-d_2))$, where $d_2>d_1\ge0$ and $\tau_m \in [d_1,d_2]$, Eq. \eqref{E_e_m_lb} can be expressed as
\begin{align}\label{error_2tap_apx}
    {\bar \varepsilon}_{m,\mathrm{lb}}^2(\boldsymbol{d}_2) &= 1-\frac{f_m^2+g_m^2-2\mathrm{sinc}(B\Delta d)f_mg_m}{1-\mathrm{sinc}^2(B\Delta d)}\ ,
\end{align}
and the optimal tap weights can be expressed as
\begin{align}\label{b_2tap_apx}
    \tilde{\boldsymbol{w}}_{\mathrm{ub},m}^\star = \frac{1}{1-\mathrm{sinc}^2(B\Delta d)}\begin{bmatrix}
        f_m-\mathrm{sinc}(B\Delta d)g_m\\
        g_m-\mathrm{sinc}(B\Delta d)f_m
    \end{bmatrix}.
\end{align}
Since $f_m$ and $g_m$ are even symmetric about $\tau_m = \frac{d_1+d_2}{2}$, we only need to focus on the value of $\tilde{w}^{\star}_{\mathrm{ub},m,1}$, which can be derived as
\begin{align}
    |\tilde{w}^{\star}_{\mathrm{ub},m,1}| \le \frac{1}{1+\mathrm{sinc}(B\Delta d)}\ .
\end{align}
Therefore, the upper bound of $\|\tilde{\boldsymbol{w}}_{\mathrm{ub},m}^\star\|_\infty$ can be derived as $\frac{1}{1+\mathrm{sinc}(1.5)}\approx 1.27$.
\end{appendices}

\bibliography{ref}

 




\vfill

\end{document}